\documentclass{article}
\usepackage{arxiv}
\usepackage[utf8]{inputenc}
\usepackage[english]{babel}
\usepackage[ruled]{algorithm2e}
\usepackage{longtable}
\usepackage{placeins}
\usepackage{amsmath}
\usepackage{amsthm}
\usepackage{graphicx}
\usepackage{color}
\usepackage{geometry}
\usepackage{adjustbox}
\usepackage{mathtools}
\usepackage{longtable}
\usepackage{placeins}
\usepackage{adjustbox}
\usepackage{booktabs}
\usepackage{placeins}
\usepackage{lineno}
\usepackage{setspace} 
\usepackage{hyperref}
\usepackage{tikz}
\usepackage{multirow}
\usepackage{siunitx}
\usepackage{caption}
\usepackage{subcaption}
\DeclareCaptionSubType * [alph]{table}
\newcommand\tab[1][1cm]{\hspace*{#1}}

\usepackage[square, numbers]{natbib}
\title{Multiple Hypothesis Hypergraph Tracking for Posture Identification in Embryonic \textit{Caenorhabditis elegans}}

\author{
 Andrew Lauziere \\
  Department of Mathematics\\
  University of Maryland, College Park\\
  College Park, MD 20742 \\
  \texttt{lauziere@umd.edu} \\
   \And
 Evan Ardiel \\
  Department of Molecular Biology at MGH\\
  Harvard Medical School\\
  Boston, MA 02115 \\
  \texttt{eardiel@yahoo.ca} \\
  \And
 Stephen Xu \\
  Laboratory of High Resolution Optical Imaging\\
  National Institutes of Health\\
  Bethesda, MD 20892 \\
  \texttt{stephen.xu.42@gmail.com} \\
  \And
Hari Shroff \\
  Laboratory of High Resolution Optical Imaging\\
  National Institutes of Health\\
  Bethesda, MD 20892 \\
  \texttt{hari.shroff@nih.gov} \\
}

\date{\today}

\begin{document}
\maketitle

\begin{abstract}
    
    \textbf{Abstract}. Current methods in multiple object tracking (MOT) rely on efficient algorithms to track large numbers of objects undergoing smooth motion. Adversarial conditions such as volatile object motion and noisy detections create a challenging tracking landscape in which established methods may yield inadequate results. These conditions often arise in applications regarding fluorescence microscopy. Trade-offs in imaging constraints emerge when balancing the observation of biological phenomena with preserving the health the specimen. Multiple hypothesis hypergraph tracking (MHHT) leverages interdependence among objects to overcome adversarial tracking conditions. The method extends traditional multiple hypothesis tracking (MHT) with hypergraphical models to quantify interdependent object motion. Both data independent and data driven hypergraphical models are proposed which can be applied to a variety of MOT applications. The motivating application concerns tracking skin cells in embryonic \textit{C. elegans}. Muscle cells cause sporadic bouts of movement in late-stage embryogenesis; however, the correlated motion in skin cells is used to improve MOT. MHHT is developed and applied to track the posture of the embryo throughout the last hours of development. Data driven hypergraphical models more accurately track posture on a held-out test embryo than methods which treat the cells as independently moving objects. Posture analysis uncovers late-stage motility defects in \textit{unc-13} mutants. 
    
\end{abstract}

\section{Introduction}

Multiple object tracking (MOT) is a fundamental task in computer vision. MOT is defined as maintaining information about a dynamic set of objects throughout a sequence of discretely sampled images; the problem is often expressed as two sequential steps: detection and association. Detection describes the process of obtaining measurements from images, while association refers to matching aforementioned measurements to unique object tracks. Accurate detections are crucial for an effective association step; however, objects moving in close proximity are difficult to track even with perfect detections. Frame-to-frame methods, such as the Global Nearest Neighbor (GNN) filter, are prone to failure when faced with these obstacles. Multiple hypothesis tracking (MHT) is a leading paradigm for the association step. The method uses detections from future frames to disambiguate challenging association decisions; it can be summarized as a multi-frame extension of the GNN approach \cite{reid_algorithm_1979,blackman_multiple_2004}. Dynamical motion models such as the Kalman Filter are used in GNN and MHT to predict object trajectories \cite{emil_rudolph_kalman_new_1960,cox_efficient_1996}. MHT with Kalman Filtering has been shown to be effective for applications in fluorescence microscopy featuring large numbers of independently moving objects with intersecting trajectories \cite{feng_multiple_2011, chenouard_multiple_2013}. 

Despite the power of MHT, the method can still be improved for emerging tracking tasks in MOT, such as scenarios featuring interdependent motion. Here we present multiple hypothesis hypergraph tracking (MHHT), a framework that extends MHT to leverage interdependent object motion to more effectively perform MOT. Relationships between groups of objects are used to give enhanced context at the data association step, yielding robust tracking in the presence of noisy detections and volatile object motion. 

\subsection{Background}

Many methods for MOT stem from the GNN approach. The basic method serves as the foundation for MHT and other algorithms such as MHT. Define $\mathbf{Z}^{(t)} = [\mathbf{z}^{(t)}_1, \mathbf{z}^{(t)}_2, \dots, \mathbf{z}^{(t)}_n]$ as the states of each object $i=1, 2, \dots, n$ at time $t=1, 2, \dots, T$. State $\mathbf{z}^{(t)}_i$ describes the center position of object $i$ at time $t$. Similarly, define $\mathbf{O}^{(t)} = [\mathbf{o}^{(t)}_1, \mathbf{o}^{(t)}_2, \dots, \mathbf{o}^{(t)}_{m^{(t)}}]$ as the set of measurements at frame $t$, indexed $j=1, 2, \dots, m^{(t)}$, $t=1,2, \dots, T$. The detection step generates sets $\mathbf{O}^{(t)}, t = 1, 2, \dots, T$, while the association step concerns using the detections to update tracks $\mathbf{Z}^{(t)}, t=1, 2, \dots, T$. 

The GNN cost matrix $\mathbf{C} \in R^{n \times (m{(t)}+n)}$ (Eq~\ref{eqn:C}) specifies costs for associating measurements $j=1, 2, \dots, m^{(t)}$ to states $i=1, 2, \dots, n$. The matrix comprises two blocks of sizes $n \times m(t)$ and $n \times n$, respectively. The first block measures the euclidean distance between track \textit{i}: $\mathbf{z}^{(t-1)}_i$ and detection \textit{j}: $\mathbf{o}^{(t)}_j$, while entries in the second block consists of costs of non association known as \textit{gates}. Each gate $d^{(t)}_i$ allows for track \textit{i} to receive no measurement at time \textit{t}. In context of the GNN, the gate specifies a radius about each track $\mathbf{z}_i^{(t-1)}$ in which measurements $\mathbf{o}_j^{(t)}$ must reside in order to associate to track \textit{i}. 

\begin{equation}
    \mathbf{C}^{(t)} = \begin{bmatrix}
\| \mathbf{z}_1^{(t-1)} - \mathbf{o}^{(t)}_1 \|_2  & \| \mathbf{z}_1^{(t-1)} - \mathbf{o}^{(t)}_2 \|_2 & ... & \| \mathbf{z}_1^{(t-1)} - \mathbf{o}^{(t)}_{m(t)} \|_2 & d_1 & \infty & \infty & ... & \infty \\
\| \mathbf{z}_2^{(t-1)} - \mathbf{o}^{(t)}_1 \|_2  & \| \mathbf{z}_2^{(t-1)} - \mathbf{o}^{(t)}_2 \|_2 & ... & \| \mathbf{z}_2^{(t-1)} - \mathbf{o}^{(t)}_{m(t)} \|_2 & \infty & d_1 & \infty & ... & \infty \\
\| \mathbf{z}_3^{(t-1)} - \mathbf{o}^{(t)}_1 \|_2  & \| \mathbf{z}_3^{(t-1)} - \mathbf{o}^{(t)}_2 \|_2 & ... & \| \mathbf{z}_3^{(t-1)} - \mathbf{o}^{(t)}_{m(t)} \|_2 & \infty & \infty & d_3 & ... & \infty \\
... & ... & ... & ... & \infty & \infty & \infty & ... & \infty \\
\| \mathbf{z}_n^{(t-1)} - \mathbf{o}^{(t)}_1 \|_2  & \| \mathbf{z}_n^{(t-1)} - \mathbf{o}^{(t)}_2 \|_2 & ... & \| \mathbf{z}_n^{(t-1)} - \mathbf{o}^{(t)}_{m(t)} \|_2 & \infty & \infty & \infty & ... & d_n \\ \end{bmatrix}  
\label{eqn:C}
\end{equation}

\begin{figure}[h!]
\centering
\includegraphics[width=.6\textwidth]{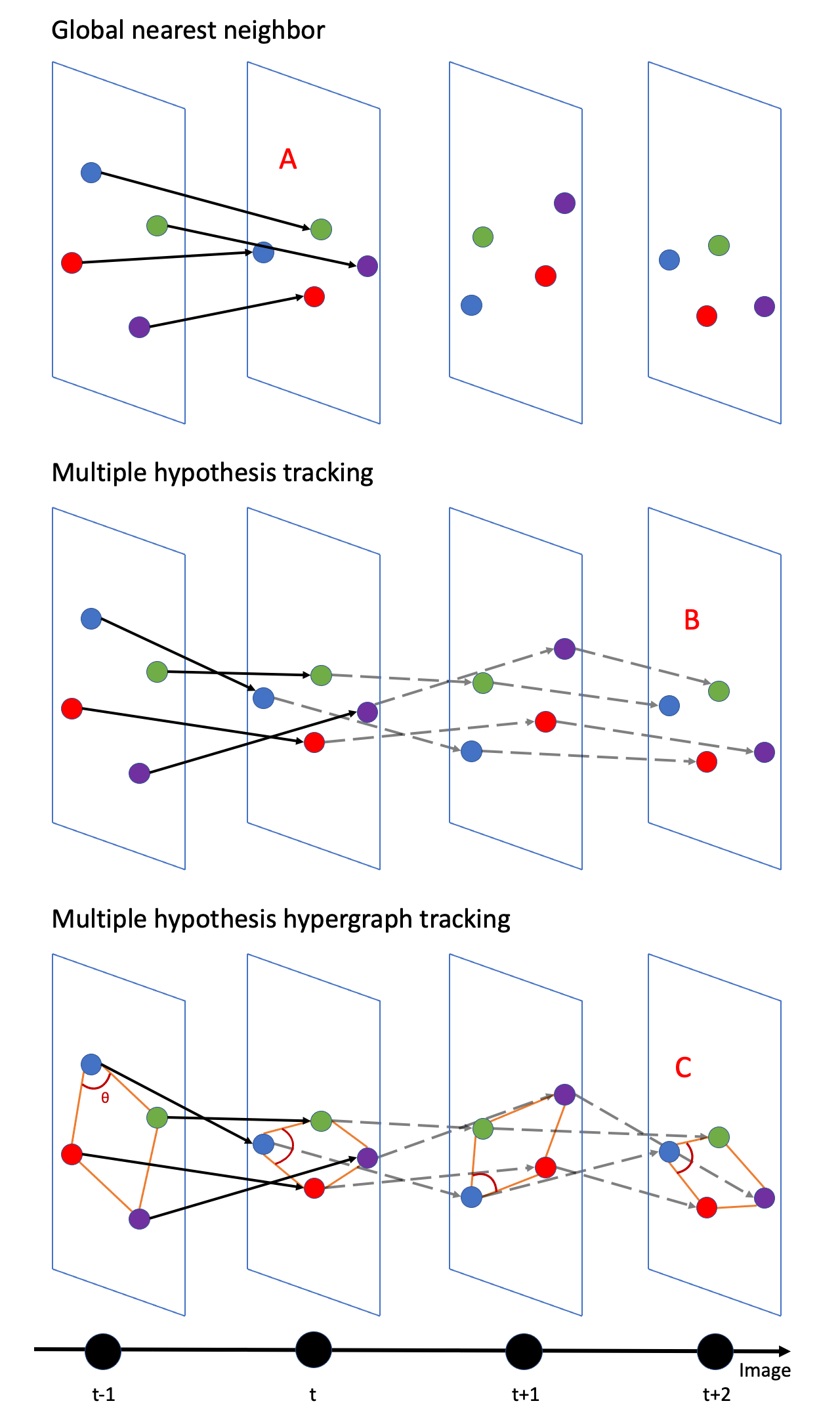}
\caption{\textbf{Interdependent object tracking can be achieved via MHHT.} Four objects (blue, red, green, purple) are observed in a sequence of images: t-1, t, t+1, and t+2. The established tracks at frame t-1 are to be continued across the next frames. Three tracking paradigms are presented: global nearest neighbor (GNN, top), multiple hypothesis tracking (MHT, middle), and multiple hypothesis hypergraph tracking (MHHT, bottom). Frame-to-frame methods such as GNN can only use information from the sequential frame (solid black arrows) to update tracks. A sudden rotation of all objects (A) may cause GNN to misidentify objects. However, Multiple Hypothesis Tracking (middle) uses information from future frames t+1 and t+2 (dashed black arrows) to disambiguate tracking decisions at frame t. Object identities can be recovered under smooth motion, but sudden large movements (B) may introduce tracking error. Both GNN and MHT perform tracking by treating objects as independent entities. MHHT (bottom) augments MHT to allow interdependent modeling of object tracks (orange lines). Correlated object movements yield additional structure such as angles (dark red). Hypergraphical modeling adds richer context than MHT or GNN, enabling accurate tracking of objects with interdependent motion (C).} 
\label{fig:MHHT_over}
\end{figure}

The cost matrix \textbf{C} defines the GNN objective, while the one-to-one and binary constraints complete the optimization problem. The resulting linear program is then solvable in polynomial time \cite{kuhn_hungarian_1955,jonker_shortest_1987}, yielding globally optimal assignments between tracks and measurements. The columns $m(t)+1, m(t)+2, \dots, m(t)+n$ correspond to a track receiving no detection update are included in the one-to-one constraints. Fig.~\ref{fig:MHHT_over}-A depicts the GNN applied to track four points. The single frame approach pairs tracks (first frame) to detections (second frame) according to which matching achieves the minimum summed distance. 

    \begin{equation}
        \begin{aligned}
                    & \text{min}
& & \sum_{i=1}^n \sum_{j=1}^{m^{(t)}} \mathbf{C}^{(t)}_{ij} x_{ij} \\
& \text{s.t.} & &  \sum_{i=1}^{n} x_{ij} = 1 \tab j = 1, \dots m^{(t)} \\
& & &  \sum_{j=1}^{m^{(t)}} x_{ij} = 1 \tab i = 1, 2, \dots n \\
& & &  x_{ij} \in \{0, 1\} \\
            \end{aligned}
            \label{eqn:GNN_LP}
    \end{equation} 
        
The GNN is the most common tool for solving the association step in MOT. The method is typically used in conjunction with physical models of object motion to a crucial step to MOT paradigms. Dynamical models of varying complexity can model first and second order liner motion, or even nonlinear trajectories.  In particular, linear dynamical models for MOT are parameterized via the Kalman Filter \cite{emil_rudolph_kalman_new_1960}. The Kalman Filter recursively updates parameters of a linear dynamical model which is used to generate a priori state predictions $\hat{\mathbf{z}_i^{(t)}}$, $i=1, 2, \dots, n$ prior to each association step. The state predictions are used in the formulation of \textbf{C}, replacing the previous frame states $\mathbf{z}_i^{(t-1)}$. Each Kalman Filter is updated after the association step, such that it better reflects object behavior. Benchmark MOT performance can be achieved by using neural networks for object detection and the GNN LP with dynamical motion modeling for association \cite{bewley_simple_2016}. 

MHT uses a ``deferred decision'' logic to make association decisions. Future detection sets are used in a probabilistic framework to disambiguate these challenging track-to-detection associations. The method requires a protocol to generate alternate association hypotheses. Murty's algorithm is the foremost method to generate the \textit{K} leading solutions to the GNN in polynomial time \cite{murty_algorithm_1968,miller_optimizing_1997}. Each association hypothesis is then a feasible full track-detection update; future detections help to delineate comparable hypotheses by individually computing likelihoods for each object. MHT can be expressed as a \textit{multidimensional assignment problem} (MAP) \cite{feng_multiple_2011, chenouard_multiple_2013}. The MAP extends the traditional LAP to more than two frames, i.e. more than two successive point-sets \cite{pierskalla_letter_1968}. Tracks at the previous frame are matched to detections in the sequential frame as a function of how those matches fit the expected object states (detections or expected locations if using a dynamical model). Detections from future frames are used to propagate tracks recursively; the matchings are subject to assignment problem constraints. The traditional MAP assumes a linear cost structure between objects, just as the MHT is expressed as well. The progression from GNN to MHT is illustrated in the second row of Fig.~\ref{fig:MHHT_over}. Track update decisions in the second frame are made according to how decisions affect future associations. The final frame (Fig.~\ref{fig:MHHT_over}-B) shows how a sudden motion can lead to incorrect associations. 

The standard linear objective (as in Eq~\ref{eqn:GNN_LP}) explicitly models associations independently, i.e. the objective states that the association of track \textit{i} to detection \textit{j} is done irrespective of \textit{i'} to \textit{j'}. While this assumption enables efficiently scaling algorithms and broad applicability, it limits the association method from using interdependencies to better describe the association problem. Graphs are abstract models that describe pair-wise relationships between objects. Vertices refer to objects themselves while edges link vertices together. Attributed vertices and edges within each the track set and detection set can then be used to describe a correspondence. For example, the distance between vertices \textit{i} and and \textit{i'} can be used when considering a potential pair-wise match to detections \textit{j} and \textit{j'}. See edges between points in the third row of Fig.~\ref{fig:MHHT_over} for an example of a graphical representation. The correspondence problem is known as \textit{graph matching}.

Hypergraphs extend the definition of a graph to include hyperedges which can specify relationships among an arbitrary number of vertices. Hypergraph matching then concerns finding an optimal vertex correspondence between pairs of attributed hypergraphs. For example, a degree three hyperedge specifies a relationship between vertices $i, i', i''$ in the track hypergraph and vertices $j, j', j''$ in the detection hypergraph. Similar to the edges between points described above, an angle is visible between the red, blue, and green points in Fig.~\ref{fig:MHHT_over}-C. Hyperedges are able to express higher degree joint relationships while edges can only express a bivariate relationship. Jointly, edges between pairs of points and angles between triplets of points assist in ensuring correct association decisions. 

Multiple hypothesis hypergraph tracking integrates the deferred decision logic of MHT with the intricate modeling capability of hypergraphs. Hypergraphical models strengthen the association step of MOT by considering an intricate representation of simultaneous track to detection pairings. Data can be used to fit the hypergraphical model, further tuning the method to a particular application. Fig.~\ref{fig:MHHT_over} depicts the progression in tracking paradigm complexity from GNN to MHHT. 

\subsection{Related Research}

Research on MOT in fluorescence microscopy has focused on single particle tracking (SPT). Particles defined as homogeneous independently moving objects are tracked under the assumption that frame-to-frame movement undergoes Brownian motion. Jaqaman et al. developed a two step approach to SPT \cite{jaqaman_robust_2008}. The first step uses an expanded GNN cost matrix to specify cutoffs for both tracks with no associated measurements and measurements that are not associated to tracks, a program similar to Eq~\ref{eqn:GNN_LP}. The second step solves a large LAP to handle linking the frame-to-frame associations, and both merge and split events throughout the image sequence. Padfield et al. combine both steps into one linear program \cite{padfield_coupled_2011}. A final image registration step is used to align successive frames with effort to clearly delineate large movements and object disappearances. The matching problem which Jaqaman et al. solved via the Hungarian algorithm is converted to a shortest path optimization problem and solved as a general linear program.

MHT based paradigms use future frame measurements, yielding a more computationally intensive but higher capacity MOT approach. Feng et al. express MHT as a multidimensional assignment problem \cite{feng_multiple_2011}. The optimization objective is then rewritten into a linear program and solved as a binary linear program. The method is compared to Jaqaman et al \cite{jaqaman_robust_2008}; MHT is more robust to noisy detections. Chenouard et al. present a probabilistic MHT framework. The MAP is again cast as a linear program to identify the optimal measurement to track association given measurements from a specified number of future frames. The program is solved via a branch-and-bound approach of the Simplex algorithm \cite{chenouard_multiple_2013}. On the other hand, Rezatofighi et al. approach joint probabilistic data association \cite{fortmann_sonar_1983}(JPDA) within an MHT paradigm. The authors propose an algorithm to rank solutions to the LAP, producing a similar output to Murty's algorithm \cite{rezatofighi_joint_2015}. The ranked solutions serve as an approximation to the JPDA association costs. The process is then integrated into an MHT scheme to robustly associate measurements to tracks by using the JPDA soft assignment criteria and the deferred logic of MHT. 

More recent methods to solve data association are inspired by the surge in popularity of neural networks. Spilger et al. demonstrate how recurrent neural networks using long-short term memory gates (LSTM) can be used to perform data association \cite{spilger_deep_2018}. The LSTM model learns assignment probabilities which then allow for the Hungarian algorithm to output a set of associations. Another recent approach uses both an LSTM and CNN for data association. Yao et al. use both networks with handcrafted and learned features to better inform the data association task \cite{yao_deep-learning_2020}. The networks are used in tandem to \textit{learn} association costs, similar to the approach of Spilger et al. \cite{spilger_deep_2018}. Neural networks are used for expressive power to better contextualize the association step. 

Euclidean distance is the underlying metric used by particle association methods \cite{chenouard_multiple_2013,rezatofighi_joint_2015, jaqaman_robust_2008}. The progression from traditional methods to recently published research is hallmarked by higher capacity models which are able to handle ambiguous association scenarios \cite{spilger_deep_2018,yao_deep-learning_2020}. MHHT builds upon the traditional MHT, but uses hypergraphical representations of object interdependencies to address challenging association decisions. Recent deep learning based approaches and proposed MHHT use data to more effectively perform data association. 

\subsection{Overview of MHHT \& Application to \textit{C. elegans}}

MHT's adaptability and intuitiveness lead to its success on MOT tasks across domains. However, challenging MOT tasks which feature interdependent object motion could better be addressed by modeling these relationships. We developed multiple hypothesis hypergraph tracking (MHHT) to flexibly integrate interdependent association models into the prominent MHT paradigm. MHHT uses graphical and hypergraphical models to contextualize relationships between objects between points. While a graphical model uses \textit{edges} to specify relationships between pairs of points, \textit{hyperedges} allow for connecting an arbitrary number of points. Graphical and hypergraphical relationships have been shown to improve the point-set matching process \cite{zhang_kergm_2019,duchenne_tensor-based_2010,lauziere_exact_2022}. Point-set matching using such complex models is known to drastically inflate computational burden \cite{sahni_p-complete_1974,lauziere_exact_2022}. However, the intricacy of the graph or hypergraph does not severely hamper computation as Murty's method is carried forward from MHT to generate hypotheses \cite{murty_algorithm_1968,miller_optimizing_1997,cox_finding_1995}. Each hypothesis constitutes a feasible full track update which is then evaluated using the posited hypergraphical model. MHHT uses Murty's algorithm in two distinct ways. The first explicitly calculates the \textit{K} best hypotheses of each hypothesis at the preceding level, i.e. generating $K^l$ hypotheses at level \textit{l} in the search tree. Hypergraphical evaluations are then conducted in a depth-first search manner to identify the minimum cost path in the tree. The second applied an optimized version of Murty's algorithm \cite{miller_optimizing_1997,cox_finding_1995} to generate the \textit{K} best hypotheses from the \textit{K} preceding hypotheses at the prior level. Results presented above were achieved using the explicit search method. 

The hypergraphical model \textit{f} yields a cost when input an association between tracks $\mathbf{Z}^{(t-1)}$ and intermediate hypotheses formed at time steps $t, t+1, \dots, t+N-1$ using detections $\mathbf{O}^{(t)}, \mathbf{O}^{(t+1)}, \dots, \mathbf{O}^{(t+N-1)}$ given the hypergraph parameters \textbf{Z}. Denote the cost $f^{(t-1,t)}$, then the hypergraphical objective between successive frames \textit{t-1} and \textit{t}. Then, the \textit{N} frame MHHT objective for the association decision at time \textit{t} can be expressed as a sum over $l=0, 1, \dots, N-1$: $\sum_{l=0}^{N-1} f^{(t+l-1,t+l)}$. The method enhances the data association step of MHT within a commonly used framework to mitigate challenges associated MOT. 

\begin{figure}
    \centering
    \includegraphics[width=\textwidth]{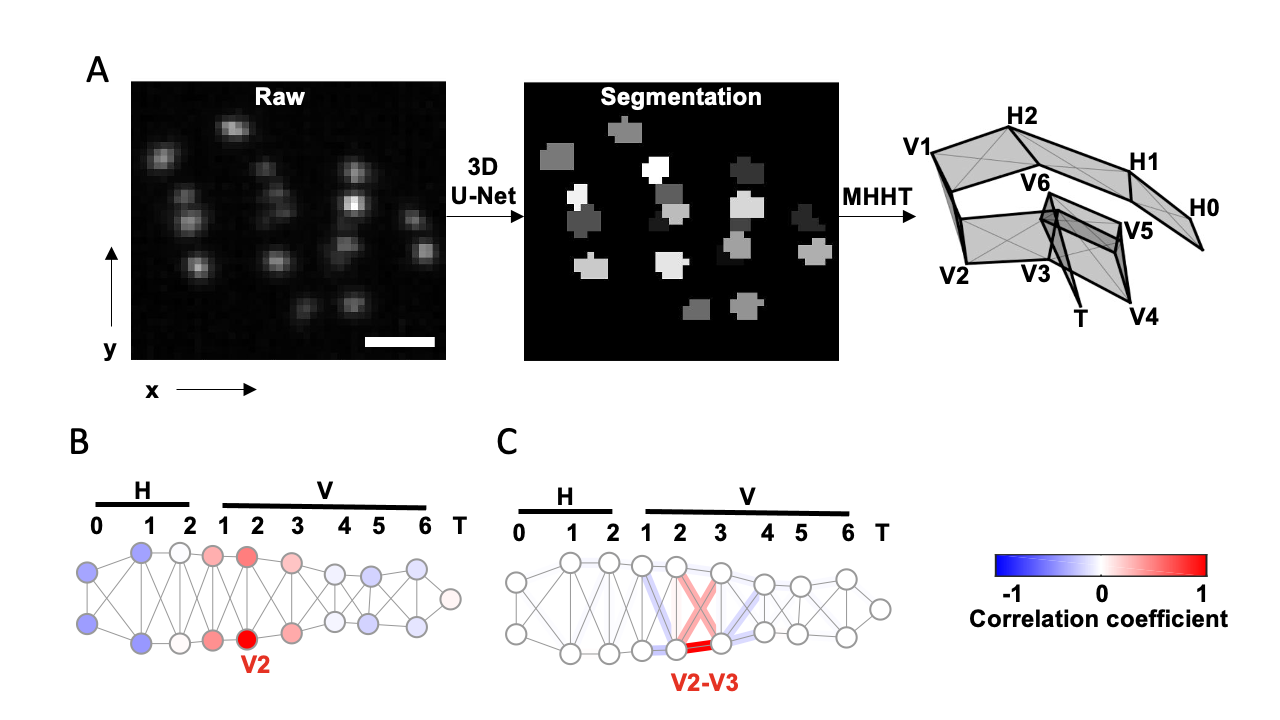}
    \caption{\textbf{Overview of MHHT applied to track posture in embryonic \textit{C. elegans}.} A) Raw image volumes are processed via a custom segmentation model to find seam cell nuclei. Scale bar 10 $\mu m$. The detection centers are then used with prior frame tracks and future detection sets to recover posture, i.e. identify \textit{all} seam cells. The process is repeated for the last hours of development, $\approx$ 54000 image volumes. B) A flattened representation of the seam cells. The cells of each pair anterior to posterior: H0, H1, ..., V5, V6, T are related to each locally other via a graph. The V2 cell's movement is positively correlated (red) with nearby cells while its movement is negatively correlated (blue) with anterior and posterior cells. C) A hypergraphical model relates edges to each other. Length variation in V2-V3 edge is positively correlated to variation of edges within that portion of the embryo while negatively correlated with edges adjacent to that portion of the embryo.}
    \label{fig:MHHT_posture_overview}
\end{figure}

MHHT was inspired by the nematode \textit{Caenorhabditis elegans} (\textit{C. elegans}), a small, free-living roundworm often studied as a model for neurodevelopment \cite{white_connectivity_1978,white_structure_1986}. The embryonic worm has a set of twenty skin cells, termed \textit{seam cells}, which act as a ``motion capture suit,'' revealing the coiled embryo's posture as it maneuvers in the eggshell. The cells run in pairs along the left and right sides of the embryo and are named in pairs: H0, H1, H2, V1, ..., V6, T, i.e. H0L and H0R comprise the H0 pair. Tracking the seam cells together constitutes recovering the coiled embryo's posture. Studying the embryo's movement throughout late-stage development yields insight into how the nervous system assumes control prior to hatching. Random bouts of muscular \textit{twitching} in late-stage development complicate posture tracking. A custom cell nucleus detection model is applied to process all $\approx$ 54000 image volumes prior to tracking (Fig.~\ref{fig:MHHT_posture_overview}-A). MHHT is applied using both data-independent and data driven hypergraphical models to perform posture tracking. The best performing models used graphical and hypergraphical relationships (Fig.~\ref{fig:MHHT_posture_overview}-B,C) to more accurately model posture. Interdependent modeling improves performance on posture tracking on images from a held-out test embryo, demonstrating the effectiveness of the method on a cutting-edge task in computational biology. 

\section{Results}

Seam cells expressing nuclear localized green fluorescent protein (GFP) were imaged by light-sheet fluorescence microscopy, which allows rapid optical sectioning with minimal photodamage \cite{wu_spatially_2013,kumar_dual-view_2014}. We recorded volumes at 3 Hz for more than 4.5 hours. Each image has $.65 \mu m^2$ resolution with a 1.2 $\mu m$ axial step size. Imaging in this manner did not appear to affect development and was not accompanied by detectable photobleaching. Consequently, we conclude that this imaging protocol did not interfere with normal embryonic development. 

To track posture, seam cell nuclei must first be accurately detected (i.e. distinguished from each other and background). Using image volumes with manually annotated seam cells, we compared the performance of several image segmentation methods. The best performance was obtained using a 3D convolutional neural network \cite{ronneberger_u-net_2015,cicek_3d_2016}. After detection, we evaluated methods for enabling comprehensive tracking of nuclear locations across all image volumes. Then, GNN and MHT approaches were compared to proposed MHHT models on posture tracking on a range of detection qualities: annotations, 3D U-Net, and IFT-Watershed \cite{falcao_image_2004,lombardot_interactive_2017}. The three methods are vary by quality of detections; the first (annotations) assuming seam cell nuclei are always detected. The 3D U-Net was the highest performing method, automatically fit on a corpus of training data. The IFT-Watershed achieved the worst performance; it was manually tuned across training image volumes. 

\begin{figure}
    \centering
    \includegraphics[width=\textwidth]{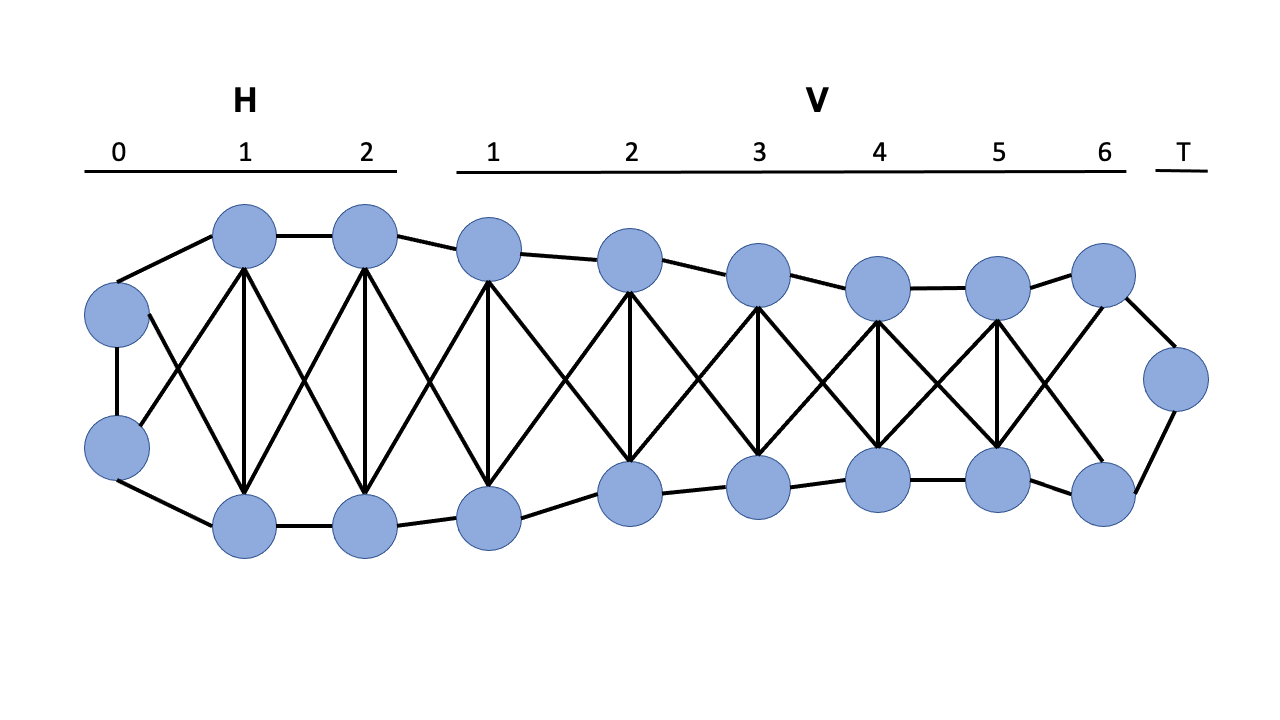}
    \caption{\textbf{The embryo graph describes interconnectivity among seam cells and serves as a basis for hypergraphical modeling.} Interdependence between seam cells is modeled via a graph. Pairs of cells are labelled anterior to posterior: H0, H1, H2, V1, ..., V5, V6, T. Physiologically close cells are linked via an edge. The graphical model underpins both the interpolation step and hypergraphical data association steps of the tracking process. Specifically, correlated movement between paired nuclei and correlated edge length variations are considered when quantifying a track update.}
    \label{fig:graph}
\end{figure}

Biological insight into \textit{C. elegans} inspired concepts of physical models to contextualize embryonic movement. The models themselves are expressed as graphs or hypergraphs. The embryo graph $G=(V,E)$ specifies a set of vertices \textit{V} representing seam cells and edges \textit{E} connecting seam cells locally. Edges appear posterior to anterior, laterally between pairs of nuclei, and diagonally between sequential pairs. The graph \textit{G} serves as the basis for graphical and hypergraphical posture tracking models. Fig.~\ref{fig:graph} depicts a the embryo graph of an uncoiled embryo. The first graphical association model, denoted \textit{Embryo}, compares changes in edge lengths frame-to-frame. Differences in the lengths of edges contribute to the cost of the track update. Annotated data were used to estimate statistics of a parametric model to further describe embryonic behavior. Two such models, \textit{Posture} and \textit{Movement}, were explored to track posture. \textit{Posture} is the data-driven enhancement of \textit{Embryo}; the hypergraphical model measures the consistency in the shape of the embryo throughout successive frames. \textit{Movement} is then a data enhanced version of the GNN, a graphical model evaluating patterned movement between nuclei. 

\subsection{Detection}

Seam cell nuclei segmentation methods were evaluated by comparing each set of detections to annotated nuclei center points at each frame. Manually labelled annotation volumes were used to generate ground truth detection sets. Voxels corresponding to nuclei were labelled $1$ while others are labelled $0$. Each independent connected component represents a single nucleus. Each volume contains the nuclei visible to the annotator; ideally there are $19$ nuclei in images from the first ~3.5 hours of post-twitch development, and $21$ when the $Q$ neuroblasts split from the $V5$ pair. The tail pair nuclei appeared close due to limited spatial resolution, and are treated as one nucleus. Some nuclei were too dim to locate, in which case they were not annotated. 

A total of 230 3D image volumes across three embryos were manually labelled using ImageJ \cite{schindelin_fiji_2012}. The binary volumes were then randomly split into training, validation, and test sets of proportions 60\%, 20\%, and 20\%, respectively.

Each proposed detection method ultimately yielded a set of detected objects. A linear program was used to match detections to annotations at each frame of the annotation set. One-to-one matching aligned annotations and detections per volume across the test set. The average precision, recall, and F1 score across test set volumes are reported for each method in Table \ref{fig:det_results}. 

\begin{table}[ht!]
    \centering
    \begin{tabular}{lrrr}
    \toprule
    \textbf{Model} &  \textbf{Precision} &  \textbf{Recall} &    \textbf{F1} \\
    \midrule
    IFT-Watershed \cite{falcao_image_2004}    &       0.81 &    0.80 &  0.80 \\
    LoG-GSF \cite{lindeberg_feature_1998, lowe_distinctive_2004}  & 0.95 & 0.90 & 0.92 \\
    Wavelet \cite{olivo-marin_extraction_2002} & 0.88 & 0.86 & 0.87 \\
    Mask-RCNN \cite{he_mask_2017}         &       0.93 &    0.89 &  0.91 \\
    3D U-Net Dice \cite{ronneberger_u-net_2015}     &       0.94 &    0.91 &  0.92 \\
    3D U-Net Dice/BCE \cite{cicek_3d_2016} &       0.95 &    0.92 &  0.93 \\
    Stardist 3D \cite{weigert_star-convex_2020}      &       0.91 &    0.88 &  0.89 \\
    \bottomrule
    \end{tabular}
    \vspace{10pt}
    \caption{\textbf{3D CNNs outperform traditional segmentation methods for seam cell nuclei detection.} Nuclear detection results on a held out test set. Large kernel 3D U-Nets yielded both the highest precision and recall. However, even the 3D U-Nets could not accurately detect all seam cell nuclei.}
    \label{fig:det_results}
\end{table}

Traditional methods for blob detection such as Watershed \cite{falcao_image_2004} and Laplacian of Gaussians (LoG) \cite{jaqaman_robust_2008} were compared to a variety of deep learning based approaches. Even recent methods such as \textit{Stardist3D} \cite{weigert_star-convex_2020} struggled to detect all seam cell nuclei. Most notable across methods is the disparity between precision and recall. While both traditional and deep learning based methods achieved comparable average precision, CNNs detected more of the dim nuclei. The latter result may be due to the CNNs' improved accuracies on voxels between close nuclei, more often returning disjoint nuclei instead of clumping multiple nuclei as detection. The adapted 3D U-Net trained on the summed Dice and BCE loss functions demonstrated the best results across the test set. 

\subsection{Tracking}

Seam cell nuclei tracking stands in contrast to typical MOT tasks in fluorescence microscopy, and thus a broader metric was used to evaluate performance. An accurate representation of the posture requires accurate tracks of \textit{all} seam cell nuclei. Incorrect associations violate the representation of the embryo and need to be corrected before cascading into further errors. The embryo also expresses several distinct behaviors throughout late-stage development, such as relative stillness, fluid movement, and twitching. Two embryos were imaged and seam cell locations were then annotated across all ($\approx$54000) volumes. The first imaged embryo was used to estimate MHHT model parameters, while performance was evaluated on the second annotated embryo, allowing for cross-validation of MHHT models. Test-set performance on the held out embryo gave insight into the generalizability of proposed models on future embryos. MHHT models were compared to two baseline methods: GNN ($K$=1, $N$=1) and MHT. MHT was conducted using MHHT with GNN costs between frames. Results presented were achieved by building explicit. 

The predicted states were compared to the annotated ground-truth coordinates at each frame. A successful posture update will align one-to-one to the annotation points with each predicted nucleus state being sufficiently close to its respective annotation point, set at 7.5 $\mu m$. Frames which failed the test with either an incorrect association or losing track of at least one nucleus marked an error which would require expert intervention, and the coordinates were reset via a correction in the next frame. Methods were compared by their ability to maintain the embryonic posture across late-stage embryogenesis throughout variations in behavior. The percentage error rate is given by the ratio of frames in which a correction is required to all $\approx$ 54000 frames on the held out test embryo.

\subsubsection{Annotations}

A series of association models $f$ were tested, increasing in complexity from the simple GNN (MHT, $K$=1,$N$=1) to the hybrid \textit{Posture-Movement} model. Varying $N$ from a single scan ($N$=1) up to $N$=5 highlighted the contribution of the deferred decision paradigm, while varying $K=1,2,\dots,5$ allowed us to employ a wider search, improving the likelihood of encountering correct posture hypotheses.

Percentage error rates on the top quartile of movement across both pre-$Q$ and post-$Q$ frames are depicted in Table \ref{fig:perf_results}. The GNN achieved a baseline 6.04\% error rate, while MHT ($K$=5, $N$=5) increases the error rate to 7.60\%, demonstrating the inability of the linear association model to distinguish competing hypotheses. However, the progression from MHT to \textit{Embryo} slightly reduced the error rate from GNN levels. The added discriminatory power of the graphical model allowed for identifying the correct posture more consistently. The data enhanced model \textit{Posture} illustrates the contribution of annotations to tracking seam cell nuclei. \textit{Posture-Movement} (\textit{PM}) yields stronger results than either alone. The combined model reduces the error rate to 3.53\%, a 42\% reduction relative to the GNN error rate. Table \ref{fig:perf_results}(a) highlights the change in error rate across models with respect to the search width $K$, holding $N$=5 constant. Table \ref{fig:perf_results}(b) shows how parameterized shape based models \textit{Posture} and \textit{PM} particularly benefit from searching deeper, illustrating an improvement in performance with tree depth $N$.

\begin{table}[h!]
\begin{subtable}[h]{0.5\textwidth}
\centering
\begin{tabular}{c||c|c|c|c|c} 
 K & \text{MHT} &  \textit{Embryo} &  \textit{Posture} &  \textit{Movement} &  \textit{PM} \\
\hline
1 &  6.04 &   6.04 &    6.04 &     6.04 &             6.04 \\
2 &  6.18 &   6.09 &    4.65 &     4.88 &             4.48 \\
3 &  6.64 &   6.03 &    4.26 &     4.54 &             4.09 \\
4 &  7.21 &   5.95 &    3.90 &     4.26 &             3.72 \\
5 &  7.60 &   5.90 &    3.71 &     4.09 &             \textbf{3.53} \\
\bottomrule
\end{tabular}
\subcaption{Varying $K$, $N$=5.}
\end{subtable}
\hfill
\begin{subtable}[h]{0.5\textwidth}
\begin{tabular}{c||c|c|c|c|c} 
 N & \text{MHT} &  \textit{Embryo} &  \textit{Posture} &  \textit{Movement} &  \textit{PM} \\
\hline
1 &  6.04 &   5.37 &    4.24 &     4.04 &             3.97 \\
2 &  6.98 &   5.53 &    4.07 &     4.06 &             3.88 \\
3 &  7.26 &   5.77 &    3.87 &     4.08 &             3.61 \\
4 &  7.41 &   5.90 &    3.82 &     4.04 &             3.58\\
5 &  7.60 &   5.90 &    3.71 &     4.09 &             \textbf{3.53}\\
\bottomrule
\end{tabular}
\subcaption{Varying $N$, $K$=5.}
\end{subtable}
\vspace{10pt}
\caption{\textbf{Data driven hypergraphical modeling better maintains posture than simpler methods.} Percentage error rates on frames in the top quartile of movement on the held out test embryo, assuming perfect detections. The combined \textit{PM} model achieves an error rate of 3.53\% ($K$=5, $N$=5), a 42\% reduction in the error rate from the baseline GNN (6.04\%). \textit{PM} also benefits from increasing $K$ and $N$, especially compared to simpler non-parametric models (MHT and \textit{Embryo}). \textbf{Bolded} entries indicate the best in class (lowest) error rates.}
\label{fig:perf_results}
\end{table}

\subsubsection{Detections}

Frames with debris and missed nuclei (i.e. imperfect detections) require evaluating more hypotheses ($K$) to find valid associations. Adapting MHHT to imperfect detections requires the introduction of gates $d^{(t)}_i$ with missing nucleus interpolation. Table \ref{fig:quart_table} compares results by detection method, 3D U-Net vs. IFT-Watershed segmentation. Baseline methods are compared to MHHT models across quartiles of frame-to-frame movement, while varying gate size and detection method; MHT and MHHT methods use $K$=25 and $N$=2. Best in-quartile error rates are in bold; highlighting the effectiveness of data driven hypergraphical models in conjunction with state of the art detection methodology. The \textit{PM} model achieved lower error rates across gate sizes and detection methods for each level of embryonic displacement, collectively lowering overall error by approximately 15\% over baseline methods. Importantly, the MHHT models also demonstrated less variance within each quartile across gates for both detection methods. In particular, the 12.5 $\mu m$ gate size posed a significant challenge for baseline methods, which were unable to distinguish between competing hypotheses. 

\begin{table}[h!]
\begin{centering}
\begin{tabular}{c|c||c|c|c|c !{\vrule width1.8pt}c |c|c|c|}
 & {\textbf{Gate} ($\mathbf{\boldsymbol{\mu} m}$)} &  \multicolumn{4}{c}{3D U-Net \cite{cicek_3d_2016}} &    \multicolumn{4}{c}{IFT-Watershed \cite{falcao_image_2004, lombardot_interactive_2017}}  \\
{} & &  GNN &    MHT &  $Embryo$ & \textit{PM} &   GNN &   MHT & $Embryo$ & \textit{PM}  \\
\midrule
\multirow{5}{*}{Q1}    & 2.5       & 3.44      & 2.94      & 2.78      & 2.82      & 3.80      & 5.09      & 3.53      & 3.68 \\
                        & 5.0       & 2.93     & 4.35      & 2.71      & \textbf{2.57}      & 5.56      & 15.87      & 3.62      & 3.91 \\
                        & 7.5       & 3.32     & 5.10     & 2.73      & 2.62      & 14.30      & 22.65      & 6.94      & 6.62 \\
                        & 10.0       & 5.60     & 5.83     & 2.93      & 2.80      & 35.37      & 30.47      & 22.12      & 20.41 \\
                        & 12.5       & 7.09     & 6.71     & 3.67      & 3.34      & 49.07      & 40.89      & 39.38      & 37.31 \\
\hline
\multirow{5}{*}{Q2}    & 2.5       & 6.00      & 5.12      & 4.72      & 4.75      & 6.68      & 7.38      & 5.87      & 6.10 \\
                        & 5.0       & 4.84     & 5.78      & 4.43      & 4.40      & 7.94      & 18.66      & 5.98      & 6.22 \\
                        & 7.5       & 4.93     & 6.70     & 4.52      & \textbf{4.31}      & 17.94      & 25.17      & 9.48      & 9.08 \\
                        & 10.0       & 6.95     & 7.47     & 4.75      & 4.49      & 38.70      & 32.57      & 25.25      & 25.25 \\
                        & 12.5       & 8.78     & 8.28     & 5.24      & 5.00      & 52.41      & 43.82      & 41.87      & 40.13 \\
\hline
\multirow{5}{*}{Q3}    & 2.5       & 10.90      & 8.40      & 8.03      & 7.86      & 11.70      & 11.58      & 9.78      & 9.67 \\
                        & 5.0       & 7.63     & 8.57      & 7.16      & 7.00      & 11.60      & 21.58      & 9.58      & 9.36 \\
                        & 7.5       & 7.80     & 9.69     & 7.17      & \textbf{6.97}      & 20.93      & 28.85      & 13.04      & 12.33 \\
                        & 10.0       & 9.91     & 10.23     & 7.21      & 7.09      & 42.25      & 36.15      & 28.00      & 26.49 \\
                        & 12.5       & 11.85     & 10.67     & 7.71      & 7.55      & 56.08      & 47.63      & 46.16      & 44.69 \\
\hline
\multirow{5}{*}{Q4}    & 2.5       & 22.68      & 19.96      & 18.70      & 18.69      & 23.86      & 22.95      & 21.71      & 21.46 \\
                        & 5.0       & 16.19     & 15.29      & 13.71      & 13.07      & 22.12      & 28.79      & 18.67      & 18.57 \\
                        & 7.5       & 14.35     & 15.81     & 12.67      & 12.19      & 29.32      & 36.20      & 21.13      & 20.95 \\
                        & 10.0       & 15.63     & 16.24     & 12.43      & \textbf{11.99}      & 49.35      & 43.23      & 35.82      & 34.63 \\
                        & 12.5       & 17.35     & 16.71     & 12.84      & 12.37      & 62.18      & 54.52      & 52.31      & 51.19 \\
\bottomrule
\end{tabular}
\vspace{10pt}
\caption{\textbf{MHHT outperforms baseline methods across levels of detection quality.} Comparing percentage error rates between baseline methods (GNN and MHT) and MHHT ($Embryo$ and \textit{PM}) when detections are imperfect. MHT and the MHHT models used $K$=25 and $N$=2. The progression in modeling capacity is compared to increasing gate size $\mu m$, embryonic movement (movement quartiles Q1-4) and detection method: 3D U-Net  \cite{cicek_3d_2016} vs. IFT-Watershed \cite{falcao_image_2004,lombardot_interactive_2017}. MHHT achieved lower error rates across all degrees of movement and detection method. In particular, \textit{PM} error rates were more robust to gate size, remaining accurate while GNN error rates increased. \textbf{Bolded} entries indicate best in class (lowest) error rates.}
\label{fig:quart_table}
\end{centering}
\end{table}

\begin{figure}[h!]
\centering
\includegraphics[width=\textwidth]{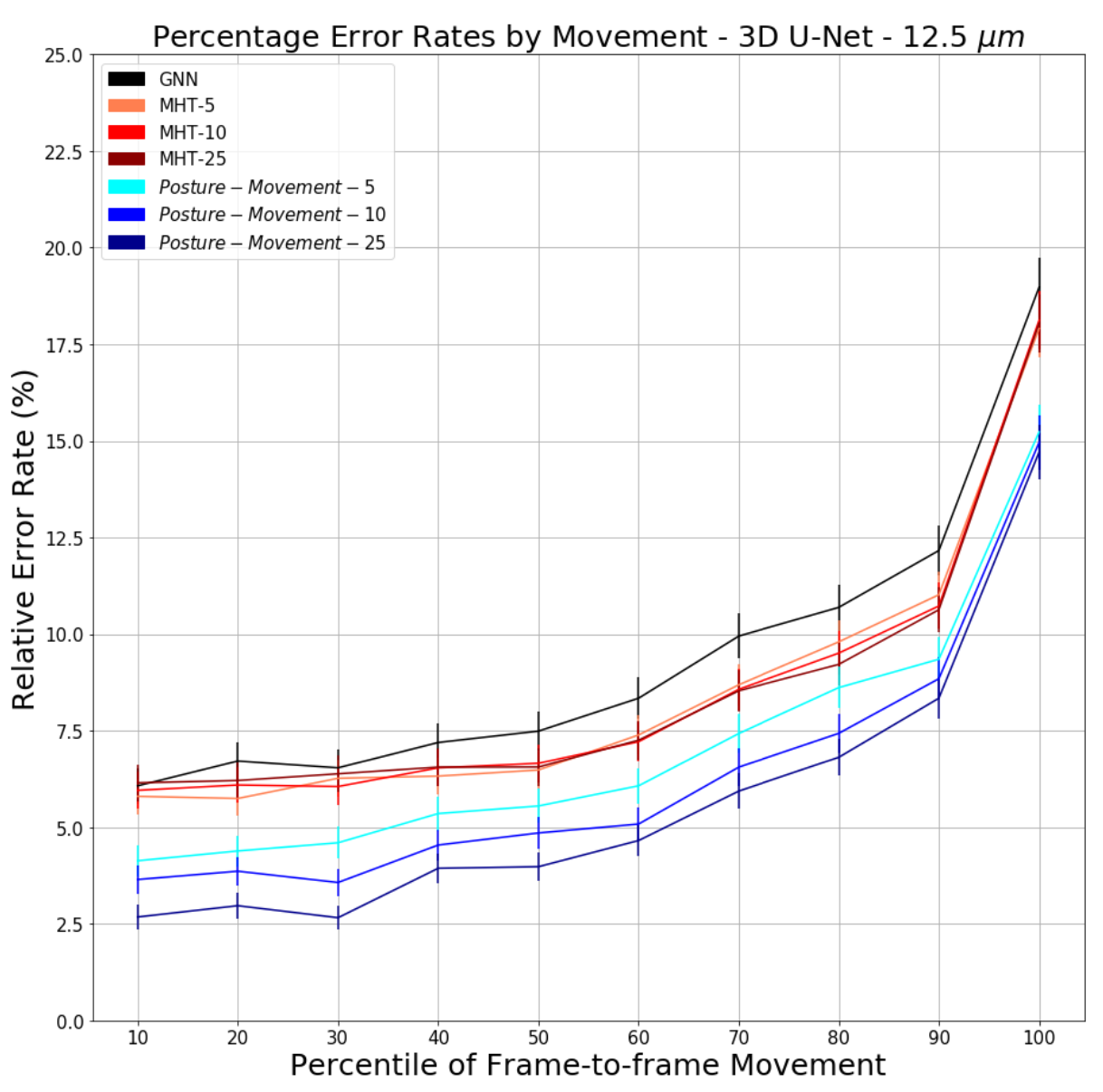}
\caption{\textbf{Hypergraphical models improve performance on challenging association decisions.} Percentage relative error rates by movement decile, focusing on the $12.5 \mu m$ gate. GNN, MHT ($K$=5, 10, 25. $N$=2), and \textit{PM} ($K$=5,10,25. $N$=2) are compared. The MHT results are clustered while the \textit{PM} results improved with increased $K$. The separation between \textit{PM} models with varying $K$ highlights the increased discriminatory power in the hypergraphical model.} 
\label{fig:quartile_det}
\end{figure}

Fig.~\ref{fig:quartile_det} highlights relative error rates on the 12.5 $\mu m$ gate across movement deciles as a function of search width $K$ (5, 10, 25) with $N=2$ on the 3D U-Net detections. \textit{PM} sees a higher marginal reduction in error rate than MHT with respect to $K$. The improvements are attributed to the enhanced discriminatory power of the hypergraphical model over unary association methods. \textit{PM} more effectively used computational resources allocated via the search width $K$ than baseline MHT to perform posture identification.

\subsection{Eigen-embryos compactly describe posture and behavioural motifs}

\begin{figure}
    \centering
    \includegraphics[width=\textwidth]{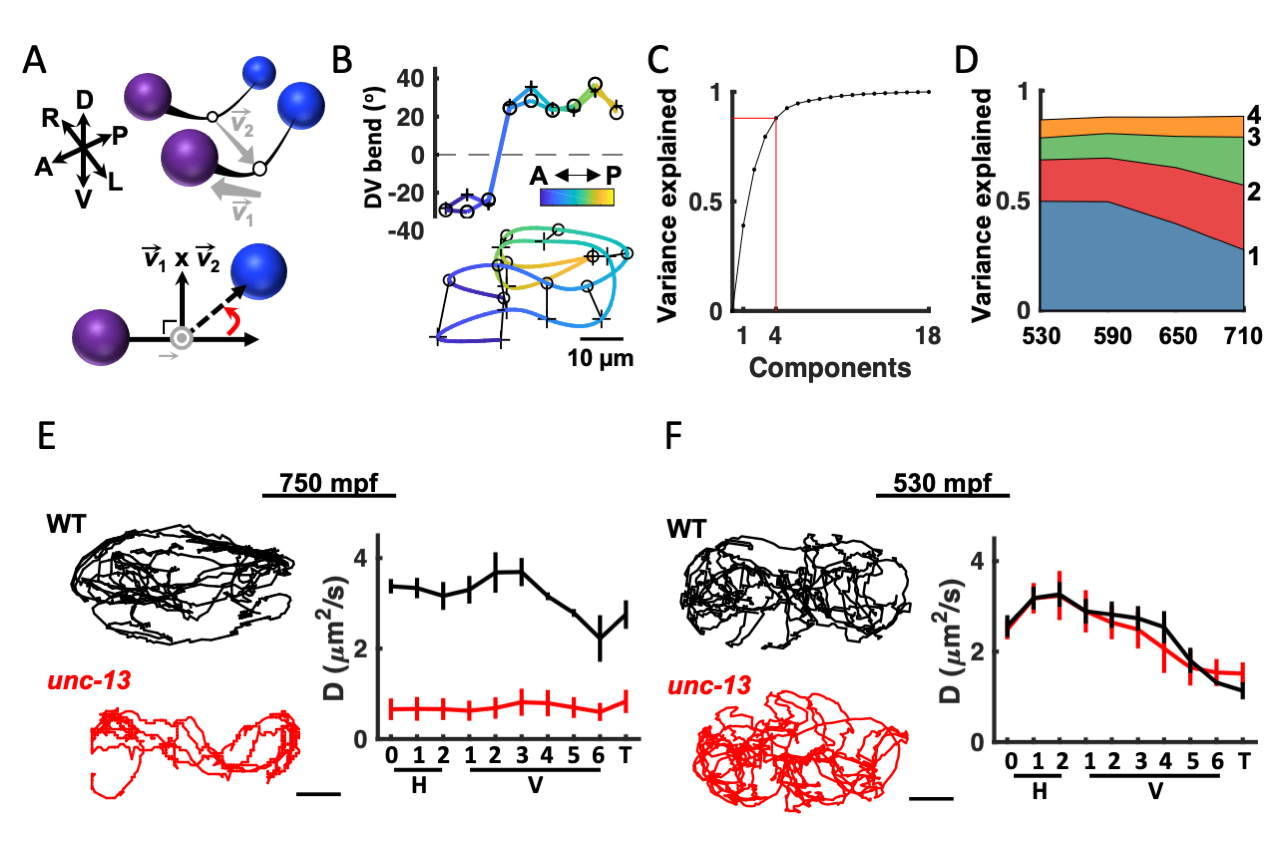}
    \caption{\textbf{Eigen-embryos compactly describe behavioral maturation.} A) A schematic illustrating how dorsoventral (DV) bends are defined. Top: Seam cell nuclei on each side of the body are fit with a natural cubic spline (black line). Vector $\Vec{v}_2$ links the midpoints between adjacent seam cell nuclei (open circles). Bottom: View looking down $\Vec{v}_2$2 to highlight a DV bend angle on one side of the body (red arrow). B) DV bend angles between all adjacent seam cells are used to define an embryo’s posture. DV bends along the left (+) and right (o) sides of an embryo (top) are plotted (top). The posture model for this embryo is shown (bottom). Position along the anteroposterior axis is indicated by the color gradient. C) The fraction of the total variance captured by reconstructing postures using 1 through all 18 principal components is plotted. D) Fraction of the total variance captured by the first 4 principal components is plotted as a function of embryo age (minutes post fertilization, mpf). E \& F) \textbf{\textit{unc-13} mutants have a late-stage motility defect.} E) Seam cell motions are compared in WT and \textit{unc-13} mutant embryos at 750 mpf (E) and 530 mpf (F). Representative 10 minute trajectories for the H1 seam cell (left) and mean diffusion coefficients for all seam cell pairs (right; mean +/- SEM) are shown. Scale bar, 10 $\mu m$. Sample sizes, 750 mpf (3 WT, 3 \textit{unc-13}); 530 mpf (3 WT, 2 \textit{unc-13}).}
    \label{fig:behavior}
\end{figure}

To describe embryo posture, we used seam cell positions to fit each side of the body with a natural cubic spline and then computed dorsoventral bending angles between adjacent seam cells (totalling 18 bend angles in each volume, Fig.~\ref{fig:behavior}-A,B). Four PCs captured approximately 88\% of variation in the 18 angles (Fig.~\ref{fig:behavior}-C). The corresponding eigenvectors of the four leading components (termed eigen-embryos) were stereotyped between animals. For example, PC1 captures ventral or dorsal coiling (i.e., all ventral or all dorsal body bends, respectively) while PC2 describes postures with opposing anterior and posterior bends. In this framework, embryonic posture can be approximated by a linear combination of eigen-embryos. The contribution of eigen-embryos shifted consistently across development, with less variance accounted for by PC1 as PC2 and PC3 gained prominence (Fig.~\ref{fig:behavior}-D). PC2 and PC3 approximate sinusoids with a phase difference of about 90$^{\circ}$ that can be combined to generate travelling waves of dorsoventral bending. The developmental shift towards PC2 and PC3 more closely approximates the adult motion pattern, where the top two PCs also describe sinusoids with a 90$^{\circ}$ phase shift \cite{stephens_dimensionality_2008}. The relatively limited dimensionality of dorsoventral bending observed in both adults and embryos is likely a consequence of muscle anatomy, with electrically coupled muscle bundles running ventrally and dorsally along the body \cite{white_structure_1986}. 

The behavior of late-stage embryos exhibits several potential signatures of neuronal control (increasingly directed movement, dorsal coiling bias, and sinuous crawling). To establish a role for synaptic signalling, we analyzed \textit{unc-13}(\textit{s69}) mutants which have a nearly complete block in synaptic vesicle fusion and (consequently) profound movement defects \cite{richmond_one_1999}. In late-stage embryos (750 mpf), \textit{unc-13} mutant movement was strongly impaired, as indicated by shorter seam cell trajectories and smaller diffusion coefficients (Fig.~\ref{fig:behavior}-E). Although their motion was severely restricted, \textit{unc-13} mutants continued subtle movements in place, suggesting that spontaneous muscle contractions persist even when synaptic transmission is blocked. By contrast, at 530 mpf seam cell diffusion coefficients for \textit{unc-13} mutants were indistinguishable from controls (Fig.~\ref{fig:behavior}-F), implying that synaptic transmission is not required for the behavior of immature embryos. See \cite{ardiel_systematic_2021} for more results and further discussion of behavioral analyses. 

\section{Discussion}

MHHT is proposed as an extension of traditional MHT to leverage complex data association functions to handle interdependent object motion in tracking applications. The method is demonstrated to be more robust to poor detections than baseline tracking paradigms while achieving lower error rates on when tracking posture in embryonic \textit{C. elegans}. Simulations across models and hyperparameter configurations demonstrate the generalizability of MHHT to similar tasks. The explicit tree search method adds discriminatory power at the cost of computation, particularly when increasing tree depth \textit{N}. On the other hand, the more typically applied version of Murty's algorithm \cite{miller_optimizing_1997, cox_finding_1995} maintains \textit{K} hypotheses at each scan. As such, the computational burden is linear in both \textit{K} and \textit{N}. Further simulations will need to be done using this method to better understand performance trade offs. 

Posture tracking in twitching embryonic \textit{C. elegans} presents itself as an emerging problem archetype in MOT. Cutting edge microscopes allow for the observation of previously intimate of biological detail. MHHT improves upon tracking outcomes across detection capabilities, using annotated data to contextualize the complete embryo frame-to-frame. The method reduces the frame-wise error rate by approximately 15\% over baseline methods on the test embryo, correctly maintaining embryonic posture and thus eliminating the need for human intervention on over 700 additional volumes, out of approximately 54000 for an imaged embryo. 

Our results suggest that embryos exhibit a stereotyped program for behavioral maturation in the final few hours before hatching, which comprises at least three phases (early flipping, an intermediate phase of reduced motility, and a late phase of mature motion). Following elongation, embryo behavior was initially dominated by flipping between all dorsal and all ventral body bends. Flipping behavior was not disrupted in \textit{unc-13} mutants, implying that it does not depend on synaptic transmission. Flipping could be mediated by intrinsic oscillatory activity in muscles or by a form of neuronal signaling that persists in the absence of UNC-13 (e.g., gap junctions or an unconventional form of synaptic vesicle exocytosis). Because flipping comprises alternating all dorsal and all ventral bends, there must be some mechanism to produce anti-correlated ventral and dorsal muscle contractions. 

Early flipping behavior is followed by a period of decreased motion. This slowdown is apparently not neuronally evoked, as it requires neither neuropeptide processing enzymes nor UNC-13. Slowing does coincide with a shift in the forces defining body morphology, from a squeeze generated by contraction of circumferential actin bundles of the epidermal cytoskeleton to containment within a tough, yet flexible extracellular cuticle \cite{priess_caenorhabditis_1986}. This structural transition could impact behavior; however, preliminary experiments (not shown) suggest that reduced motion is a consequence of decreased muscle activity rather than a structural constraint limiting motion. A shift from cytoskeletal to exoskeletal control of body shape is likely reiterated at each larval molt and could contribute to molt-associated lethargus quiescence \cite{raizen_lethargus_2008}. 

Following the inactive phase, late-stage embryos exhibit a mature pattern of motion. Mature motion comprises sinusoidal crawling, prolonged bouts of forward and reverse motion, and a rhythmic pattern of brief quiescent bouts. All of these features are grossly disrupted in \textit{unc-13} mutants, implying that they are driven by synaptic circuits. Our prior study suggests that bouts of directed motion are mediated by the forward and reversal locomotion circuits that operate post-hatching \cite{ardiel_visualizing_2017}. 

\section{Methods}

\subsection{Multiple Hypothesis Hypergraph Tracking}

\subsubsection{Hypergraphical Association}

The established MHT paradigm \cite{blackman_multiple_2004,cox_efficient_1996} relies upon a linear data association step to evaluate hypotheses. Successive associations across future frames ideally disambiguate the correct track update among sampled hypotheses. A hypergraphical model can better evaluate the sampled hypotheses by accounting for interdependent object motion. An association function \textit{f} measures the dissimilarity between established tracks $\mathbf{Z}^{(t-1)}$ and a hypothesized state update $\hat{\mathbf{Z}}^{(t)})$. The state estimate of the $K$ sampled hypotheses $\hat{\mathbf{Z}}^{(t,k)}$ which minimizes the association function $f$ is chosen as the joint state update, $\mathbf{Z}^{(t)}$. The method is most effective when a linear program can identify competitive state updates, but cannot accurately discern between candidates solely using the linear program costs. 

A graph  $G=(V,E)$ specifies a set of edges \textit{E} connecting objects locally. Hypergraphical models allow for further evaluation of a sampled hypothesis under some further function of this edge set \textit{E}. Hypotheses evaluated at time \textit{t}: $\hat{\mathbf{g}}^{(t)} \vcentcolon = \mathbf{g}(\hat{\mathbf{Z}}^{(t)}; E)$ describe the hypothesis according to predicted states at time $t$. Frame-to-frame differences in the attributed representations are assumed multivariate Gaussian: $(\hat{\mathbf{g}}^{(t)} - \mathbf{g}^{(t-1)}) \sim \mathcal{N}(\mathbf{0}, \mathbf{\Sigma})$. The Mahalanobis distance $f$ is used to evaluate a hypothesized state representation:

\begin{equation*}
    f(\hat{\mathbf{g}}^{(t)}, \mathbf{g}^{(t-1)}; \hat{\mathbf{\Sigma}}^{-1}) = \sqrt{(\hat{\mathbf{g}}^{(t)} - \mathbf{g}^{(t-1)})^{'}\hat{\mathbf{\Sigma}}^{-1}(\hat{\mathbf{g}}^{(t)} - \mathbf{g}^{(t-1)})}
\end{equation*}

The covariance matrix $\mathbf{\Sigma}$ is estimated from a corpus of annotated data. States of all \textit{n} objects from frames $t=1, 2, \dots, T$ are used as pairs $\{(\mathbf{Z}^{(1)}, \mathbf{Z}^{(2)}), (\mathbf{Z}^{(2)}, \mathbf{Z}^{(3)}), \dots, (\mathbf{Z}^{(T-1)}, \mathbf{Z}^{(T)})\}$ to estimate frame-to-frame variation in pairs of hyperedge differences: $\{(\mathbf{g}^{(1)}, \mathbf{g}^{(2)}), (\mathbf{g}^{(2)}, \mathbf{g}^{(3)}), \dots, (\mathbf{g}^{(T-1)}, \mathbf{g}^{(T)})\}$. Define $\Bar{\mathbf{g}} = \frac{\sum_{t=1}^T \mathbf{g}^{(t+1)} - \mathbf{g}^{(t)}}{T-1}$. Then, the covariance matrix is estimated: 

\begin{equation*}
    \hat{\mathbf{\Sigma}} \vcentcolon = \frac{1}{T-1} \sum_{t=1}^T (\mathbf{g}^{(t+1)} - \mathbf{g}^{(t)} - \bar{\mathbf{g}})(\mathbf{g}^{(t+1)} - \mathbf{g}^{(t)} - \bar{\mathbf{g}})^{'}
\end{equation*}

\subsubsection{Graphical Interpolation}

Independently moving objects cannot provide insight into states of objects with missing measurements. However, specified dependencies  between objects can be used to more precisely update tracks that do not receive a measurement at frame $t$. Our proposed graphical interpolation uses states $\mathbf{Z}^{(t-1)}$ and a hypothesis $\varphi^{(t,k)}$ with measurements $\mathbf{O}^{(t)}$ to complete the state update with predictions $\bar{\mathbf{z}}^{(t)}_i$, $i=1, 2, \dots, n$. The intermediate state update of hypothesis $k$: $\hat{\mathbf{Z}}^(t,k)$ can then be written in terms of associated measurements $\mathbf{O}^{(t)}$ and predicted object positions $\bar{\mathbf{Z}}^{(t)}$:

    \begin{equation}
        \begin{aligned}
            \hat{\mathbf{z}}^{(t,k)}_{i} = \begin{cases}
            \bar{\mathbf{z}}^{(t,k)}_i & \varphi_k^{(t)}=0 \\
           \mathbf{o}^{(t)}_j & \varphi_k^{(t)}=j
            
                \end{cases}
        \end{aligned}
        \label{eqn:hypo}
    \end{equation}

A graphical model specifying connectivity between objects serves as the basis for point interpolation. Each hypothesis $\varphi_k^{(t)}$ describes two disjoint sets of objects: objects which receive a measurement at $t$ ($\varphi_k^{(t)} \neq 0$) and objects that do not receive a measurement at $t$ ($\varphi_k^{(t)}=0$). Designate the two disjoint sets $D^{(t,k)}$ and $U^{(t,k)}$, detected vertices and undetected vertices, respectively. Edges in which one element $u$ is missing the other $v$ is detected are used to predict the state of the missing vertex $u$. In the scenario in which for all edges connecting a missing vertex $u$, $v$ is also a missing vertex the prior frame position is used for interpolation. Assume each undetected object $u \in U^{(t,k)}$ is a component of some edges in which the complementing vertex is detected: $\Tilde{D}_u^{(t,k)} = \{v \in D^{(t,k)}: (u,v) \in E\} \subset U^{(t,k)}$, $n^{(t,k)}_u \vcentcolon = |\Tilde{D}_u^{(t,k)}|$. The graphically interpolated states are expressed:

    \begin{equation}
        \bar{\mathbf{z}}^{(t,k)}_u = \frac{\sum_{v \in \Tilde{D}_u^{(t,k)}} [\hat{\mathbf{z}}^{(t,k)}_{v} - (\mathbf{z}^{(t-1)}_v - \mathbf{z}^{(t-1)}_u)]}{n^{(t,k)}_u}
        \label{eqn:interp}
    \end{equation}

The estimated position of a missing object is the average of predicted positions under the edge set $E$. Inconsistencies in the complete state update will be penalized by hypergraphical association models. The method is extendable to hypergraphical or other forms of state interpolation involving object features derived from the image. 

\begin{figure}
    \centering
    \includegraphics[width=\textwidth]{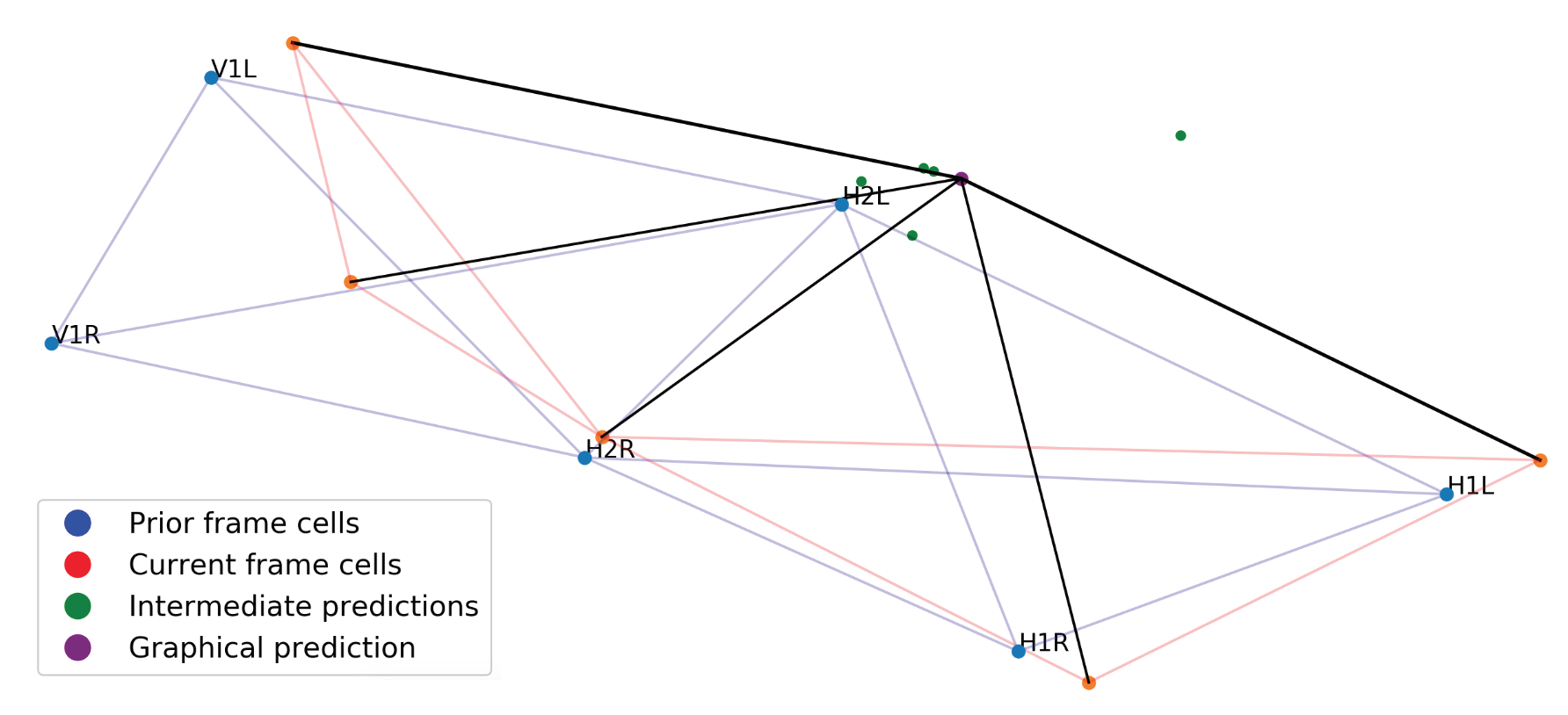}
    \caption{\textbf{Graphical interpolation leverages hypothesized associations to yield a more robust full track update.} Graphical interpolation is applied to posture tracking via the embryo graph. The blue graph represents an anterior part of the posture of the image volume show in Fig.~\ref{fig:graph}. The red graph arises from a posture hypothesis in the next frame in which H2L is missing. Black edges comprising identified relevant nuclei are used in conjunction with the prior frame graph to predict the position of H2L. The green points are intermediate predictions with the purple point being the final prediction, an average of individually predicted black points via Eq~\ref{eqn:interp}.}
    \label{fig:interp}
\end{figure}

Fig.~\ref{fig:interp} illustrates an example of graphical interpolation applied to posture tracking. An anterior portion of Fig.~\ref{fig:graph} is depicted with H2L missing. Five nuclei give insight into H2L's position: H1L, H1R, H2R, V1L, and V1R. The prior volume points (blue) and the identified current volume points (red) yield intermediate predictions (green). The final prediction (purple) is an average of all five intermediate predictions. Encoded domain knowledge of \textit{C. elegans} body structure via the graph is used to prune hypotheses which result in physically invalid postures.

\subsubsection{Algorithm}

MHHT adapts the established hypothesis oriented MHT proposed by Cox and Hingorani \cite{cox_efficient_1996} to include graphical interpolation and further evaluation of the sampled hypotheses. In summary, Murty's algorithm identifies leading solutions to the data association problem proposed in Equation \ref{eqn:GNN_LP}; states that are not directly updated undergo graphical interpolation when possible, and a hypergraphical association function $f$ measures the cost of the complete state updates $\hat{\mathbf{Z}}^{(t,k)}$. This process is iterated recursively across multiple future frames to better inform the state update at time $t$. The parameter $N$ dictates how many future detection sets are considered. The deferred decision logic considers the expansion from states $\hat{\mathbf{Z}}^{(t,k)}$ to $\hat{\mathbf{Z}}^{(t+1,k)}$, $\hat{\mathbf{Z}}^{(t+2,k)}$, \dots, $\hat{\mathbf{Z}}^{(t+N-1,k)}$ in the update at time $t$.

\begin{figure}
    \centering
    \includegraphics[width=\textwidth]{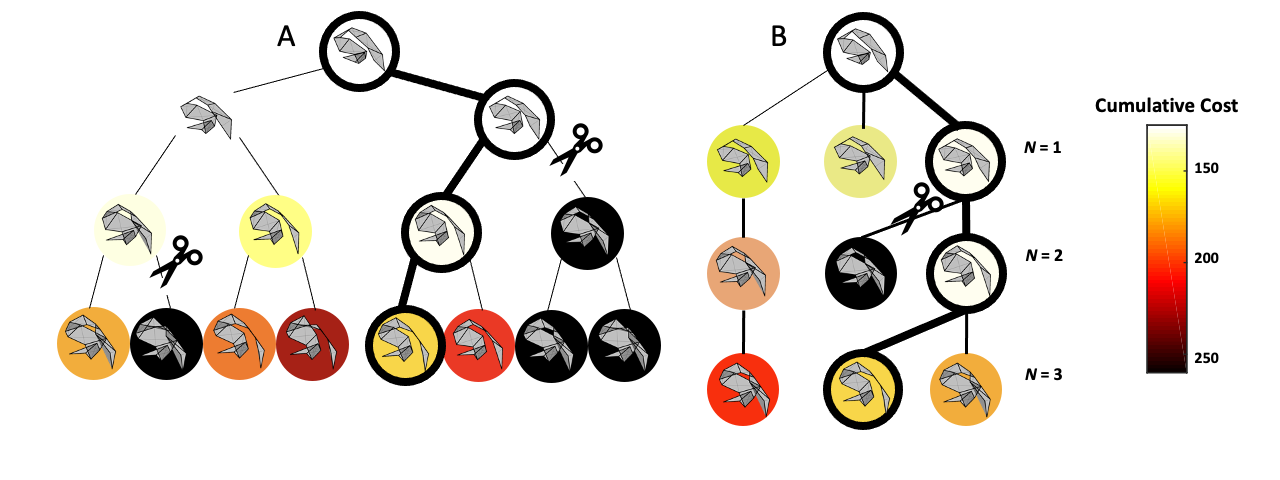}
    \caption{\textbf{MHHT search trees constructed using two different methods.} Murty's algorithm applied in two different manners to perform posture tracking. A) The traditional implementation of Murty's algorithm will generate the \textit{K} best hypotheses from each preceding hypothesis. The exponentially growing search tree yields a more thorough exploration of hypotheses. On the other hand, a more typically used version of Murty's algorithm will generate the \textit{K} best hypotheses from the \textit{K} preceding hypotheses. Worst case computation is reduced from exponential in \textit{N} to linear in \textit{N}. Bolded circles highlight the cost-minimizing path; scissors denote a pruned hypothesis due to Murty's algorithm returning a biologically infeasible posture.}
    \label{fig:MHHT_trees}
\end{figure}

The recursion can be executed in two different ways. The first method forms an exponentially growing search tree in which paths from the initial state $\mathbf{Z}^{(t-1)}$ expand into $K$ solutions. Each of the $K$ solutions following interpolation yields a hypothesized state $\hat{\mathbf{Z}}^{(t,k)}$, initiating a recursion, adding to the cost of association at frame $t$. The search follows in a depth-first search manner with respect to time. The first complete posture sequence across \textit{N} frames stands as the minimum cost hypothesis. The search continues with pruning to find the cost minimizing hypothesis accessible in the search tree. There are $\mathcal{O}(K^N)$ hypotheses are evaluated in the worst-case in the explicit tree search. Fig.~\ref{fig:MHHT_trees}-A depicts the explicit tree search method with \textit{K}=2 and \textit{N}=3. Murty's algorithm \cite{miller_optimizing_1997} is applied at each hypothesis to generate \textit{K} hypotheses at the next time point. The cost-minimizing path is bolded, with the right hypothesis at \textit{N}=1 being the chosen full-track update. Scissors indicate a pruned path due to an infeasible posture position. Darker colors (white to dark red) indicate an increasing cumulative cost as tracks are added to the search path.

The second method follows that of Miller, Stone and Cox \cite{miller_optimizing_1997} and Cox and Miller \cite{cox_finding_1995} in which the \textit{K} best hypotheses of the \textit{K} preceding hypotheses are found in one call of Murty's algorithm \cite{murty_algorithm_1968}. This second method is used in leading MHT applications \cite{cox_efficient_1996}. The trade off is in losing the ability to compare \textit{all} $K^2$ hypotheses from a previous subproblem as opposed to only the \textit{K} leading hypotheses (according to the linear model). There are in total $NK$ hypotheses evaluated using this method. Fig.~\ref{fig:MHHT_trees}-B shows the optimized version of Murty's algorithm typically applied in MHT \cite{cox_efficient_1996} with \textit{K}=3 and \textit{N}=3. Again, the bolded path identifies the cost-minimizing path. However, not all $K^l$ hypotheses at level \textit{l} (but for \textit{l}=1) are evaluated using the hypergraphical model. As such, a suboptimal path may be returned in exchange for reduced computation. 

The introduction of graphical modeling requires a fixed structure among tracked objects, with the $n$ objects being set a priori. New objects require updating the graphical models for interpolation and association, and will be tracked until manually removed from the object set. MHHT is designed to track a set number of objects with specified interdependencies despite noisy detections and large bouts of coherent motion. 

\subsection{Posture Tracking in Embryonic \textit{C. elegans}}

\subsubsection{Overview}

Posture is defined as the complete state identification of \textit{all} seam cells, which approximates the shape of the coiled embryo. Seam cell nuclei tracking is achieved via a \textit{detect and track} paradigm. The image volumes are first processed in batch via a 3D convolutional neural network (CNN) to detect nuclei. The detections are used to track seam cells throughout the image sequence. Tracking is achieved via our proposed method: multiple hypothesis hypergraph tracking (MHHT). 

\subsubsection{Seam Cell Nuclei Detection} \label{Detection}

\begin{figure}[h]
\centering
\includegraphics[width=\textwidth]{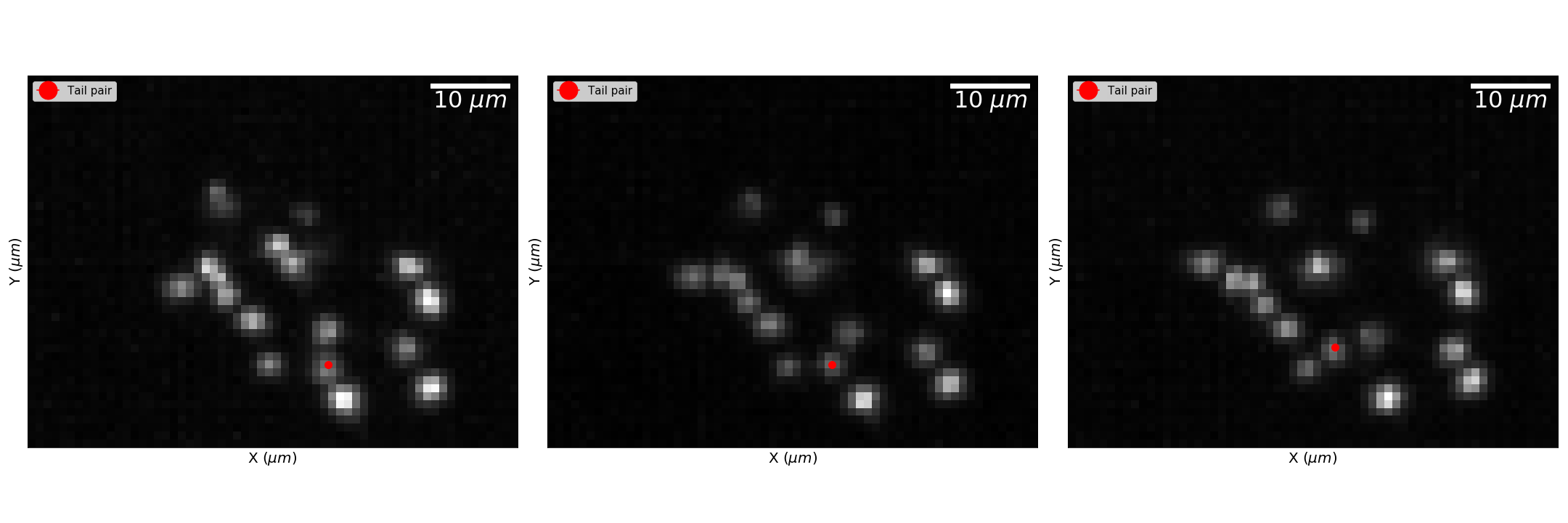}
\caption{\textbf{Low spatial resolution, high temporal resolution volumetric imaging allows observation of late-stage behavior.}Three sequential XY maximum intensity projections of images captured at approximately 2 hours before hatching, imaged at 3 Hz. Low illumination dose permits long-term imaging of rapid movements in the embryo without obvious phototoxicity. The apparent size and intensity of seam cell nuclei fluctuate throughout imaging, but are largely homogeneous and are indistinguishable based on appearance alone. Red dots are placed on nuclei in the tail pair.} 
\label{fig:max_projs}
\end{figure}

Accurate seam cell nuclei detection is especially challenging for two distinct reasons. First, nuclei frequently appear dim to the degree that an expert has to infer their positions from the more visible adjacent nuclei. The second issue arises due to the spatial resolution of the image volumes. The imaged embryo will often depict two seam cell nuclei so close to each other that they appear as one larger nucleus. Fig.~\ref{fig:max_projs} depicts three XY maximum intensity projections from sequential image volumes. The shapes and intensities of fluorescently labelled seam cell nuclei vary throughout imaging. The tail nuclei are smaller than other nuclei and much closer than nuclei of other pairs. The limited spatial resolution results in the tail nuclei appearing merged as one nucleus. Red dots are placed on seam cell nuclei in the tail pair in each maximum intensity projection. 

\begin{figure}[h]
\centering
\includegraphics[width=\textwidth]{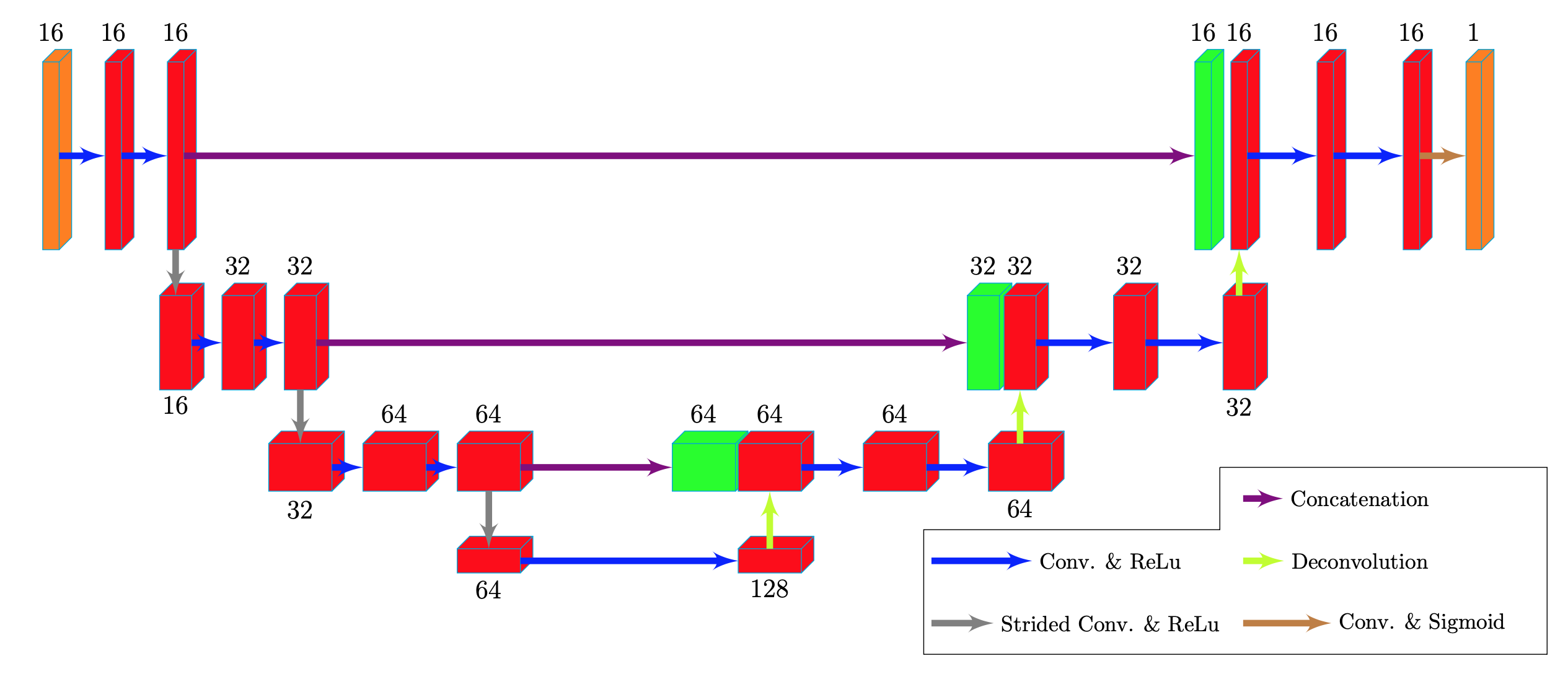}
\caption{\textbf{A 3D U-Net is trained to perform semantic segmentation on image volumes.} The input (orange) is a full 3D image volume. Successive convolutions (blue arrows) and strided convolutions (gray arrows) encode information. Deconvolution layers parametrically upsample input layers towards the input size. The number of filters in each layer (1, 16, 32, 64, 128) is displayed. The output is (orange) is the same size as the input, each voxel value $y_{ijk} \in [0,1]$ representing the likelihood of being part of a nucleus.} 
\label{fig:Unet}
\end{figure}

Seam cell nuclei detection is achieved by a convolutional neural network (CNN). A 3D U-Net style architecture is employed to perform semantic segmentation on the \textit{C. elegans} embryo image volumes \cite{ronneberger_u-net_2015, cicek_3d_2016}. Fig.~\ref{fig:Unet} depicts the model architecture. The established 3D U-Net is augmented to use a size $(5,5,5)$ kernel, extending the field of view at each layer. Strided 3D convolution layers downsample by a factor of two across lateral dimensions while preserving axial resolution. The limited axial resolution encodes more information per planar image due to the explicit downsampling occurring during imaging. The number of filters in each layer doubles from $16$ to $32, 64, 128$ when downsampling. 

The loss function is a uniformly weighted average of the binary cross-entropy and negative dice coefficient. The cross-entropy portion prioritizes accurate prediction of challenging individual voxels between close nuclei, but may produce noisy predictions. On the other hand, the dice coefficient prioritizes structural similarity in clusters of voxels constituting nuclei, but is known to consolidate sigmoid outputs at extreme values 0 and 1, resulting in close nuclei being labelled as one supervoxel \cite{scherr_best_2018}. Denote $\mathbf{y} \in \{0,1\}^{X \times Y \times Z}$ and $\hat{\mathbf{y}} \in [0,1]^{X \times Y \times Z}$  as the binary ground truth and sigmoid output tensors respectively. The loss can then be written:


\begin{equation*}
\mathcal{L}(\mathbf{y}, \hat{\mathbf{y}}) = \underbrace{- \sum_{i=1}^{X} \sum_{j=1}^{Y} \sum_{k=1}^{Z} [y_{ijk}\log(\hat{y}_{ijk}) + (1-y_{ijk})(\log(1-\hat{y}_{ijk})]}_{\text{Cross Entropy}} + \underbrace{\frac{-2\sum_{i=1}^{X} \sum_{j=1}^{Y} \sum_{k=1}^{Z} y_{ijk}\hat{y}_{ijk}}{\sum_{i=1}^{X} \sum_{j=1}^{Y} \sum_{k=1}^{Z} (y_{ijk}^2 + \hat{y}^2_{ijk})}}_{\text{Dice Coefficient}}
\end{equation*}

The model is trained via the Adam optimizer with an initial learning rate of $0.00075$ \cite{kingma_adam_2017}. The resulting image volume contains values $\hat{y}_{ijk} \in [0,1]$. The output is thresholded and then independent connected components are returned as predicted instances of seam cell nuclei. The centroids of each supervoxel serve as the detection set for each image volume. 

Imperfect detections due to low spatial resolution contribute to a challenging MOT task. Detection methods often yield debris and fail to detect all nuclei due to either dimness or clumping adjacent nuclei together. The proposed 3D U-Net \cite{cicek_3d_2016} outperforms both traditional methods: Imaging Forest Transform Watershed \cite{falcao_image_2004}, Laplacian of Gaussian \cite{lindeberg_feature_1998, lowe_distinctive_2004} and a Wavelet method \cite{olivo-marin_extraction_2002}, and modern CNN based approaches: Mask-RCNN \cite{he_mask_2017}, and Stardist 3D \cite{weigert_star-convex_2020} in detecting seam cell nuclei. Descriptions and implementations of detection methods can be found in the appendix (section \ref{det_review}). 

\subsubsection{Posture Tracking}

Posture tracking is achieved using MHHT with a proposed hybrid graphical and hypergraphical model. The explicit tree search method of Murty's algorithm is used to achieve the lowest cost hypotheses at expense of computation. Disjoint body segments are formed by sequential pairs of seam cells. Incorrect associations may yield a graphical representation in which body segments intersect with each other. The M\"{o}ller-Trombore algorithm is used to estimate intersection among body segments and prune invalid hypotheses \cite{moller_fast_1997}. 

\paragraph{\textit{Embryo} \& \textit{Posture}} 

The \textit{Embryo} and \textit{Posture} models penalizes associations which distort the embryo's shape. Accurate seam cell nuclei state updates will adhere to the physical confines of the embryo, even during sudden twitches. Edges illustrated in Fig.~\ref{fig:graph} vary in length as the states update frame to frame. Denote the edges $E = [e_1, e_2, \dots, e_M]$, where edge $e_j = (u_j,v_j)$ describes a relationship between nuclei $u_j$ and $v_j$. Each vector $\mathbf{e}^{(t)}_j = \mathbf{z}^{(t)}_{u_j} - \mathbf{z}^{(t)}_{v_j}$ describes the chord connecting states of nuclei $u_j$ and $v_j$ at time $t$. Then, the vector $\mathbf{E}^{(t)} = [\| \mathbf{e}^{(t)}_1 \|_2, \| \mathbf{e}^{(t)}_2 \|_2, \dots, \| \mathbf{e}^{(t)}_M \|_2]$ describes lengths of these chords. Differences in chord lengths between frames $\mathbf{E}^{(t)}$ - $\mathbf{E}^{(t-1)} \in R^{M \times 1}$ form the basis of the association cost. Then, the \textit{Embryo} model is defined:

\begin{equation}
    f_{E}(\hat{\mathbf{Z}}^{(t)}, \mathbf{Z}^{(t-1)}) = \sqrt{(\hat{\mathbf{E}}^{(t)} - \mathbf{E}^{(t-1)})^{'}\mathbf{I}(\hat{\mathbf{E}}^{(t)} - \mathbf{E}^{(t-1)})} = \sum_{j=1}^M \sqrt{(\hat{\mathbf{E}}^{(t)} - \mathbf{E}^{(t-1)})^2}
\end{equation}

Annotated data are used to estimate covariances between the $M$ differences across state updates. The resulting covariance matrix $\hat{\Sigma}_P$ scales differences in chord lengths among the $M$ chords present in $G$:

\begin{equation}
    f_{P}(\hat{\mathbf{Z}}^{(t)}, \mathbf{Z}^{(t-1)}; \hat{\mathbf{\Sigma}}^{-1}_P) = \sqrt{(\hat{\mathbf{E}}^{(t)} - \mathbf{E}^{(t-1)})^{'}\hat{\mathbf{\Sigma}}^{-1}_P(\hat{\mathbf{E}}^{(t)} - \mathbf{E}^{(t-1)})}
\end{equation}

Both \textit{Embryo} and \textit{Posture} are further characterized by the unary costs specified by $\mathbf{C}$ (Equation \ref{eqn:C}) and the evaluated hypothesis $\varphi^{(t)}$: $\sum_{i=1}^n \mathbf{C}_{i,\varphi^{(t)}}$. 

\paragraph{\textit{Movement}}

\textit{Movement} extends the traditional GNN cost to penalize \textit{unnatural} movement between states. The distance between states at $t-1$ and $t$, $\sum_{i=1}^n \| \mathbf{z}^{(t)}_i - \mathbf{z}^{(t-1)}_i \|_2$, can be scaled by the inverse covariance matrix describing motion between pairs of nuclei. The states $\mathbf{z}^{(t)}_{i} = [x^{(t)}_{i}, y^{(t)}_{i}, z^{(t)}_{i}]$ and $\mathbf{z}^{(t-1)}_{j} = [x^{(t-1)}_{j}, y^{(t-1)}_{j}, z^{(t-1)}_{j}]$ can be expressed as element-wise differences: 

    \begin{equation}
        \mathbf{Z}^{(t)} - \mathbf{Z}^{(t-1)} = \begin{bmatrix} \mathbf{z}^{(t)}_{1} \\ \mathbf{z}^{(t)}_{1} \\ \vdots \\ \mathbf{z}^{(t)}_{n} \end{bmatrix}  - \begin{bmatrix} \mathbf{z}^{(t-1)}_1 \\ \mathbf{z}^{(t-1)}_2 \\ \vdots \\ \mathbf{z}^{(t-1)}_n \end{bmatrix} =  \begin{bmatrix} x^{(t)}_1 \\ y^{(t)}_1 \\ z^{(t)}_1 \\ x^{(t)}_2 \\ y^{(t)}_2 \\ z^{(t)}_2 \\ \vdots \\ x^{(t)}_n \\ y^{(t)}_n \\ z^{(t)}_n \end{bmatrix} -  \begin{bmatrix} x^{(t-1)}_1 \\ y^{(t-1)}_1 \\ z^{(t-1)}_1 \\ x^{(t-1)}_2 \\ y^{(t-1)}_2 \\ z^{(t-1)}_2 \\ \vdots \\ x^{(t-1)}_n \\ y^{(t-1)}_n \\ z^{(t-1)}_n \end{bmatrix} =  \begin{bmatrix} x^{(t)}_1 - x^{(t-1)}_1 \\ y^{(t)}_1 - y^{(t-1)}_1 \\ z^{(t)}_1 - z^{(t-1)}_1 \\ x^{(t)}_2 - x^{(t-1)}_2 \\ y^{(t)}_2 - y^{(t-1)}_2 \\ z^{(t)}_2 - z^{(t-1)}_2 \\ \vdots \\ x^{(t)}_n - x^{(t-1)}_n \\ y^{(t)}_n - y^{(t-1)}_n \\ z^{(t)}_n - z^{(t-1)}_n \end{bmatrix} \in R^{3n \times 1}
    \end{equation}

Each pair of nuclei has an estimable $3 \times 3$ covariance matrix specifying the relationship between movement along each axis. The resulting block $3n \times 3n$ covariance matrix, $\mathbf{\Sigma}_M$ then scales the difference between states:

\begin{equation}
    f_{M}(\hat{\mathbf{Z}}^{(t)}, \mathbf{Z}^{(t-1)}; \hat{\mathbf{\Sigma}}^{-1}_M) = \sqrt{(\hat{\mathbf{Z}}^{(t)} - \mathbf{Z}^{(t-1)})^{'}\hat{\mathbf{\Sigma}}^{-1}_M(\hat{\mathbf{Z}}^{(t)} - \mathbf{Z}^{(t-1)})}
\end{equation}

The \textit{Posture} and \textit{Movement} models are combined additively to produce the \textit{Posture-Movement} (\textit{PM}) model.

\subsubsection{An Interface for Posture Tracking}

We developed our detect and track method with the goal of extracting seam cell coordinates over time, thereby allowing detailed analysis of embryonic behavior. The image volumes are processed in batch, yielding the complete detection set. An interface is developed in Python to perform posture tracking in embryonic \textit{C. elegans} via \textit{MHHT}. The web based interface is freely available: \url{https://github.com/lauziere/MHHT}. Track correction is performed in \textit{MIPAV}: Medical Imaging, Processing, and Visualization \cite{mcauliffe_medical_2001, christensen_untwisting_2015}. MIPAV is available: \url{https://mipav.cit.nih.gov/}. 

\subsubsection{Creating Annotations}

Image volumes were processed in batch via the large kernel 3D U-Net in section \ref{Detection}. Tracking was achieved via a gated GNN. An overlaid graphical representation (Fig.~\ref{fig:graph}) serves as visual cue for correctly updating posture. The user is prompted to verify if the posture is correct; the program generates a file structure that MIPAV can recognize and outputs the necessary parameters to view the data if the tracks need to be edited. The process is applied recursively frame-to-frame throughout the image sequence until hatching. 

\section{Acknowledgements}

This research is supported by the Laboratory of High Resolution Optical Imaging within the National Institute of Biomedical Imaging and Bioengineering at the National Institutes of Health. This work used the computational resources of the NIH HPC Biowulf cluster. (http://hpc.nih.gov). We  thank Mr. Brandon Harvey for providing annotation data and Dr. Ghadi Salem for early advice on nuclear segmentation strategies. Andrew Lauziere's contribution to this research was supported in part by NSF award DGE-1632976. The code and data that support these studies are available at \url{https://github.com/lauziere/MHHT}. 

\clearpage


\clearpage

\section{Appendix}

\subsection{Object Detection in Fluorescence Microscopy} \label{det_review}

Image segmentation aims to partition an image into disjoint regions formed by sets of connected pixels. Image segmentation methods can be split into two groups: instance segmentation and semantic segmentation. Semantic segmentation assigns a class label to each pixel in an image (or voxel in an image volume), while instance segmentation identifies clusters of pixels defining an instance of an object within an image. Semantic segmentation methods serve as an initial step towards identifying instances of objects of interest. Postprocessing routines such as a connected components analysis will cluster pixels together to yield disjoint objects. 

Traditional methods in segmentation for fluorescence microscopy leverage the blob like profile of imaged objects. A blob describes a cluster of pixels belonging to one object. A Gaussian intensity profile usually describes each fluorescent blob like object well; roughly spherical or ellipsoidal objects exhibit peak intensity near the center and lower intensity towards the edges. Locating homogeneous objects such as cells, cell nuclei, or particles is then equivalent to detecting blobs in an image. Both traditional and modern strategies are evaluated to perform seam cell nuclei detection. Traditional methods rely on homogeneous blob like structures while modern deep learning based methods use a corpus of training data with a highly parameterized (millions of parameters) graphical model. Table \ref{fig:det_ref_table} lists each method and its implementation for seam cell nuclei detection. 

\begin{table}[ht!]
    \centering
    \begin{tabular}{c|c|c}
    \toprule
    Method &  Citation & Implementation \\
    \midrule
    IFT-Watershed        &   \cite{falcao_image_2004}    &     \cite{lombardot_interactive_2017} \\
    LoG-GSF                  &   \cite{lindeberg_feature_1998, lowe_distinctive_2004} &     \cite{tinevez_trackmate_2017} \\
    Wavelet         &      \cite{olivo-marin_extraction_2002} &  \cite{de_chaumont_icy_2012} \\
    Mask-RCNN  &    \cite{he_mask_2017} &   \cite{waleed_mask_2017} \\
    3D U-Net &  \cite{cicek_3d_2016}     &   \cite{chollet_keras_2021}    \\
    Stardist 3D & \cite{weigert_star-convex_2020}     &   \cite{weigart_stardist_2021} \\
    \bottomrule
    \end{tabular}
    \vspace{10pt}
    \caption{Multiple strategies are evaluated to perform seam cell nuclei detection. Both traditional methods (IFT-Watershed, LoG-GSF, Wavelet) and modern deep learning based methods (Mask-RCNN, 3D U-Net, Stardist 3D) are listed with implementation citations.}
    \label{fig:det_ref_table}
\end{table}

Three traditional segmentation approaches were applied: Image Foresting Transform Watershed (IFT-Watershed) \cite{falcao_image_2004}, Laplacian of Gaussian with Gaussian shape fitting (LoG-GSF) \cite{lindeberg_feature_1998, lowe_distinctive_2004}, and a Wavelet based method \cite{olivo-marin_extraction_2002}. The IFT-Watershed algorithm uses a bottom-up approach to merge maxima from a Euclidean distance transform into disjoint regions \cite{falcao_image_2004}. LoG-GSF first applies the Laplacian of Gaussian, a staple method for blob detection \cite{lindeberg_feature_1998}. Resulting spots are filtered according to a quadratic fitting scheme \cite{lowe_distinctive_2004}. The Wavelet method uses a wavelet transform decomposition across multiple scales to identify bright spots \cite{olivo-marin_extraction_2002}. These methods use few parameters and are able to capture homogeneous bright spots effectively. 

Neural networks are parametric graphical models that leverage large amounts of annotated data to \textit{learn} a compositional functional relationship between inputs and outputs. Convolutional neural networks (CNNs) are a specific type of neural network better able to process both planar and volumetric image data; these models yield the best results in biological image segmentation \cite{ronneberger_u-net_2015, cicek_3d_2016, weigert_star-convex_2020}. Fully convolutional networks (FCNs) are a subclass of CNNs that output an image or volumetric image the same shape as the input, known as the encoder-decoder network architecture \cite{long_fully_2015}. Image features are extracted from the image, and iteratively downsampled and processed throughout the network to expand field of view while learning more abstract representations. The U-Net demonstrated how FCNs could revolutionize semantic segmentation in microscopy \cite{ronneberger_u-net_2015}. The RCNN extended the FCN to perform instance segmentation with the region proposal network of \citet{he_mask_2017}. However, the Mask-RCNN is currently only able to process images, not volumes. Semantic segmentation is possible on image volumes via volumetric convolutions. The 3D U-Net demonstrated the effectiveness of stacked images as context for segmenting microscopy images \cite{cicek_3d_2016}. More recently, Stardist 3D combines elements of a volumetric FCN, such as the 3D U-Net, but with a focus on identifying disjoint objects in fluorescence microscopy \cite{weigert_star-convex_2020}. The key contribution of Stardist 3D is a processing algorithm which inscribes convex polyhedra into detected blobs with the goal of separating close or touching objects.

\end{document}